\pdfoutput=1
\documentclass[aps,prd,amsmath,floats,floatfix, twocolumn,
superscriptaddress,nofootinbib,showpacs]{revtex4-1}

\usepackage[T1]{fontenc}
\usepackage[utf8]{inputenc}
\usepackage{lmodern}
\usepackage{tensor}
\usepackage{verbatim}
\usepackage[caption=false]{subfig}
\usepackage{wasysym}

\usepackage[dvipsnames, usenames]{xcolor}
\definecolor{linkcolor}{rgb}{0.0,0.3,0.5}
\usepackage[hypertexnames=false, unicode, colorlinks=true, linkcolor=linkcolor,
citecolor=linkcolor, filecolor=linkcolor,urlcolor=linkcolor,
pdfusetitle]{hyperref}

\usepackage[all]{hypcap}
\usepackage{graphicx}
\usepackage{xspace}
\usepackage{amssymb}
\usepackage[normalem]{ulem} 
\usepackage{bm} 
\usepackage{enumitem,amssymb}

\usepackage{microtype}

\usepackage[english]{babel}
\usepackage{blindtext}
\usepackage{tikz}
\usetikzlibrary{decorations.pathmorphing}
\usetikzlibrary{decorations.markings}
\usepackage{orcidlink}

\graphicspath{%
  {figs/}%
}

\DeclareMathAlphabet{\mathpzc}{OT1}{pzc}{m}{it}

\newcommand{\hertz}{\Psi_{\rm ORG}}
\newcommand{\hertzbar}{\bar{\Psi}_{\rm ORG}}
\newcommand{\etal}{\textit{et al.\ }}

\newlist{todolist}{itemize}{2}
\setlist[todolist]{label=$\square$}

\allowdisplaybreaks

\begin{document}

\title{The excitation of quadratic quasinormal modes for Kerr black holes}

\newcommand{\Cornell}{\affiliation{Cornell Center for Astrophysics
    and Planetary Science, Cornell University, Ithaca, New York 14853, USA}}
\newcommand\CornellPhys{\affiliation{Department of Physics, Cornell
    University, Ithaca, New York 14853, USA}}
\newcommand\Caltech{\affiliation{TAPIR 350-17, California Institute of
    Technology, 1200 E California Boulevard, Pasadena, CA 91125, USA}}
\newcommand{\AEI}{\affiliation{Max Planck Institute for Gravitational Physics
    (Albert Einstein Institute), Am M\"uhlenberg 1, Potsdam 14476, Germany}} %
\newcommand{\UMassD}{\affiliation{Department of Mathematics,
    Center for Scientific Computing and Visualization Research,
    University of Massachusetts, Dartmouth, MA 02747, USA}}
\newcommand\Olemiss{\affiliation{Department of Physics and Astronomy,
    The University of Mississippi, University, MS 38677, USA}}
\newcommand{\Bham}{\affiliation{School of Physics and Astronomy and Institute
    for Gravitational Wave Astronomy, University of Birmingham, Birmingham, B15
    2TT, UK}}
\newcommand{\Perimeter}{\affiliation{Perimeter Institute for Theoretical Physics, Waterloo, ON N2L2Y5, Canada}}
\newcommand{\UGuelph}{\affiliation{University of Guelph, Guelph, Ontario N1G 2W1, Canada}}
\newcommand{\Tsinghua}{\affiliation{Department of Astronomy, Tsinghua University, Beijing 100084, China}}

\author{Sizheng Ma\orcidlink{0000-0002-4645-453X}}
\email{sma2@perimeterinstitute.ca}
\Perimeter

\author{Huan Yang\orcidlink{0000-0002-9965-3030}}
\email{hyangdoa@tsinghua.edu.cn}
\Tsinghua
\Perimeter
\UGuelph

\hypersetup{pdfauthor={Ma et al.}}

\date{\today}

\begin{abstract}
The excitation of quadratic quasinormal modes is an important nonlinear phenomenon for a Kerr black hole ringing at a specific linear mode. The amplitude of this second-order effect is proportional to the square of the linear mode amplitude, with the ratio being linked to the nature of the Kerr black hole. Focusing on the linear $(l=m=2,n=0)$ mode, we compute the dependency of the ratio on the dimensionless spin of the black hole, ranging up to 0.99, with the method applicable for more general mode couplings. Our calculation makes use of the frequency-domain, second-order Teukolsky equation, which involves two essential steps (a) analytically reconstructing the metric through the Chrzanowski-Cohen-Kegeles approach and (b) numerically solving the second-order Teukolsky equation using the shooting method along a complex contour. We find that the spin dependence of the ratio shows a strong correlation with the angular overlap between parent and child modes, providing qualitative insights into the origin of the dependence. Depending on the nature of the angular overlap, the ratio decreases with spin in scenarios such as the channel $(l=m=2,n=0)\times(l=m=2,n=0)\to(l=m=4)$ or increases in situations like $(l=m=2,n=0)\times(l=m=2,n=0)\to(l=5,m=4)$. For both cases, the ratios do not vanish in the extremal limit. Our studies may offer insights into the search for quadratic quasinormal modes from numerical relativity and gravitational wave detections. As a byproduct, we find that the Weyl scalars can be concisely expressed with the Hertz potential.
\end{abstract}

\maketitle


\section{Introduction}
\label{sec:introduction}
The final stage of a binary black hole (BH) coalescence process is described by a deformed Kerr BH ringing down and relaxing to its final state.
Depending on the starting time of the analysis, the ringdown process may be well-described by the BH perturbation theory, and the resultant GW waveform is consistent with a superposition of quasinormal modes (QNMs) \cite{Buonanno:2006ui,Berti:2007fi,Giesler:2019uxc}. Leveraging the fact that the QNM frequencies are uniquely determined by the mass and spin of the remnant Kerr BH, as motivated by the no-hair theorem \cite{Carter:NoHair}, the idea of BH spectroscopy \cite{Berti:2005ys,Berti:2016lat,Yang:2017zxs} has been developed and implemented: by comparing theoretical predictions of QNM frequencies with observational data from the LIGO-Virgo-KAGRA network \cite{TheLIGOScientific:2016src,LIGOScientific:2020tif,Abbott:2020mjq,LIGOScientific:2021sio,Berti:2018vdi,Carullo:2019flw,Isi:2019aib,Siegel:2023lxl,Capano:2021etf,Finch:2022ynt,Cotesta:2022pci,Ma:2023vvr,Isi:2019aib,Wang:2023xsy}, one can test General Relativity and constrain the parameters of modified gravity theories.


Moving deeper within the merger phase, it is natural to expect nonlinear effects of General Relativity should play more important roles both in the dynamics of the spacetime and in the observed data. 
In previous literature, various kinds of nonlinear effects have been discussed to different extents. Gleiser \etal \cite{Gleiser:1995gx} initiated the analysis with second-order Regge-Wheeler-Zerilli formalism. The second-order Teukolsky formalism was later developed by Campanelli and Lousto \cite{Campanelli:1998jv}.  Zlochower \etal \cite{Zlochower:2003yh} investigated mode coupling in the scattering processes by BHs. Yang \etal studied parametric coupling between modes to motivate a transient instability of near-extremal Kerr BHs \cite{Yang:2014tla,Yang:2015jja}. More recently the nonlinear excitation of modes in dynamical BHs with varying mass and spin was analyzed in Refs.~\cite{Sberna:2021eui,Redondo-Yuste:2023ipg}. In addition,  quadratic QNMs have been identified from numerical-relativity simulations both in waveforms at future null infinity \cite{Ma:2022wpv,Mitman:2022qdl,Cheung:2022rbm} and in multipole moments of dynamical horizons \cite{Khera:2023oyf}. Then a related interesting question is:  {\it{how do the quadratic QNMs reflect the nature of Kerr BHs?}} To answer this question, it is preferable to analyze quadratic QNMs for different BH spins, and possibly compare them with predictions from modified theories of gravity. This work targets the first part of this task, i.e., the predictions within General Relativity.

The dynamics of quadratic QNMs are captured by second-order perturbation theory \cite{Gleiser:1995gx,Campanelli:1998jv,Nakano:2007cj,Ioka:2007ak,Loutrel:2020wbw,Ripley:2020xby,Lagos:2022otp,Spiers:2023cip}. These modes originate from the interaction between two linear parent QNMs. Consequently, the complex frequencies of the quadratic QNMs emerge as a linear combination of the frequencies of the individual parent modes and their respective complex conjugates. Meanwhile, the quadratic amplitudes are proportional to the product of the linear amplitudes, with the ratio being called the ``excitation factor''. The factor is an intrinsic parameter of the Kerr spacetime which characterizes the level of difficulty to excite the corresponding quadratic modes. In the context of Schwarzschild BHs, Nakano and Ioka \cite{Nakano:2007cj,Ioka:2007ak} used the Regge-Wheeler-Zerilli formalism to compute the factor for the channel $(l=2,m=2,n=0)^2 \to (l=4,m=4)$, where $(l,m)$ are angular indices and $n$ stands for the overtone index. The Wentzel–Kramers–Brillouin \cite{Perrone:2023jzq} and uniform \cite{Bucciotti:2023ets} approximations were also applied to provide some analytical insights for the excitation factor of Schwarzschild BHs. As for Kerr BHs, the Kerr/CFT correspondence \cite{Kehagias:2023ctr} was used to estimate the excitation factors in the near-extremal limit. In addition, two recent studies \cite{Redondo-Yuste:2023seq,Zhu:2024rej} analyzed the problem with a purely numerical perspective: fitting the amplitudes of child and parent modes obtained from scattering experiments. 

In this work, we provide a systematic description of the quadratic QNMs by computing their excitation factors with a semi-analytical approach. We achieve this goal by solving the frequency-domain, second-order Teukolsky equation. Our methodology comprises two key steps: (a) analytically reconstructing the metric through the method given by Chrzanowski-Cohen-Kegeles (CCK) \cite{1975PhRvD..11.2042C,COHEN19755,1979PhRvD..19.1641K} and (b) numerically solving the second-order Teukolsky equation using the shooting method along a complex contour. We implement this method for specific mode couplings and compare the results with previous literature.

This paper is organized as follows. In Sec.~\ref{sec:perturbation_review}, we briefly review the linear perturbation of Kerr BHs. Section \ref{sec:metric_reconstruction} focuses on the CCK method. We then use this method to construct an analytical source for the second-order Teukolsky equation in Sec.~\ref{sec:second_order_Teukolsky_equation}. Our numerical approach to solving the Teukolsky equation in the complex $r$-domain is introduced in Sec.~\ref{sec:contour}. It is then applied to address the problem of quadratic QNMs in Sec.~\ref{sec:results}. Finally, we summarize the results in Sec.~\ref{sec:conclusion}.

Throughout this paper we fix the mass of Kerr BHs at $M=1$, therefore the symbol $a$ is used to denote their dimensionless spins. Our metric signature is $(+, -, -, -)$. Complex conjugates are represented by overlines.


\section{Black-hole perturbations via Newman-Penrose formalism}
\label{sec:perturbation_review}
In this section, we provide a brief overview of the linear perturbation of Kerr BHs, with more detailed references given in \cite{Kokkotas:1999bd,Berti:2009kk,chandrasekhar1998mathematical}. In particular, we review the derivative operators and the Teukolsky equation used in later calculations. Using Boyer-Lindquist coordinates, the Kerr metric is expressed as follows
\begin{align}
    &ds^2=\frac{\Delta}{\Sigma} (dt-a\sin^2\theta d\phi)^2-\frac{\sin^2\theta}{\Sigma}\left[(r^2+a^2)d\phi-adt\right]^2\notag \\
    &-\frac{\Sigma}{\Delta}dr^2-\Sigma d\theta^2,
\end{align}
where 
\begin{align}
    \Sigma=r^2+a^2\cos^2\theta, \quad \Delta=r^2-2r+a^2. \notag 
\end{align}
The spacetime is accompanied with an event horizon at $r_+=1+\sqrt{1-a^2}$, and a Cauchy horizon at $r_-=1-\sqrt{1-a^2}$.
A Kerr BH has two principal null directions, described by the Kinnersley tetrad
\begin{equation}
\label{eq:Kinnersley}
   \begin{aligned}
    &\bm{l}=\frac{1}{\Delta}\left(r^2+a^2,\Delta,0,a\right), \\
    &\bm{n}=\frac{1}{2\Sigma}\left(r^2+a^2,-\Delta,0,a\right), \\
    & \bm{m} = \frac{1}{\sqrt{2}\Gamma}(ia\sin\theta, 0, 1, i\csc\theta),
\end{aligned} 
\end{equation}
with $\Gamma=r+ia\cos\theta$. 

The perturbation theory of Kerr is based on the Newman-Penrose (NP) formalism, where spin coefficients and Weyl scalars serve as fundamental quantities. Using the Kinnersley tetrad, these quantities are given by:
\begin{align}
    &\kappa^{(0)}=\sigma^{(0)}=\lambda^{(0)}=\nu^{(0)}=\epsilon^{(0)}=0, \quad \rho^{(0)}=-\frac{1}{\bar{\Gamma}}, \notag \\
    &\beta^{(0)}=\frac{\cot\theta}{2^{3/2}\Gamma},\quad \pi^{(0)}=\frac{ia\sin\theta}{2^{1/2}\bar{\Gamma}^2}, \quad \tau^{(0)}=-\frac{ia\sin\theta}{2^{1/2}\Sigma}, \notag  \\ 
    &\mu^{(0)}=-\frac{\Delta}{2\Sigma\bar{\Gamma}}, \quad \gamma^{(0)}=\mu^{(0)}+\frac{r-1}{2\Sigma}, \quad \alpha^{(0)}=\pi^{(0)}-\bar{\beta}^{(0)}, \label{eq:NP_spin_coefficients}
\end{align}
and $\Psi_4^{(0)}=\Psi_3^{(0)}=\Psi_1^{(0)}=\Psi_0^{(0)}=0, \Psi_2^{(0)}=-\bar{\Gamma}^{-3}$. The overline operation $\bar{(...)}$ denotes the complex conjugation.

\subsection{Derivative operators}
There are three primary derivative operators in the NP formalism
\begin{align}
    &\bm{D}^{(0)}=l^\mu\nabla_\mu=\partial_r+\frac{r^2+a^2}{\Delta}\partial_t+\frac{a}{\Delta}\partial_\phi,\notag \\
    &\bm{\Delta}^{(0)}=n^\mu\nabla_\mu=-\frac{\Delta}{2\Sigma}\left(\partial_r-\frac{r^2+a^2}{\Delta}\partial_t-\frac{a}{\Delta}\partial_\phi\right),\notag \\
    &\bm{\delta}^{(0)}=m^\mu\nabla_\mu=\frac{1}{\sqrt{2}\Gamma}\left(\partial_\theta+i\csc\theta\partial_\phi+ia\sin\theta\partial_t\right). \notag 
\end{align}
Note that the derivative operator $\bm{\Delta}^{(0)}$ is not to be confused with the scalar function $\Delta$. To facilitate our upcoming derivation, it is advantageous to follow Chandrasekhar \cite{chandrasekhar1998mathematical} (see Chapter 8) and define operators
\begin{equation}
  \begin{aligned}
    &\mathcal{D}_n=\partial_r+\frac{(r^2+a^2)}{\Delta}\partial_t+\frac{a}{\Delta}\partial_\phi+2n\frac{r-1}{\Delta}, \\
    &\mathcal{D}_n^\dagger=\partial_r-\frac{(r^2+a^2)}{\Delta}\partial_t-\frac{a}{\Delta}\partial_\phi+2n\frac{r-1}{\Delta}, \\
    &\mathcal{L}_n^\dagger=\partial_\theta+ i\csc\theta\partial_\phi+ia \sin\theta\partial_t+n\cot\theta,\\
    &\mathcal{L}_n=\partial_\theta- i\csc\theta\partial_\phi-ia \sin\theta\partial_t+n\cot\theta.
\end{aligned}  
\end{equation}
Here $n$ is typically an integer. We can see that $\mathcal{D}$ and $\mathcal{D}^\dagger$ do not have $\theta-$dependence, whereas $\mathcal{L}$ and $\mathcal{L}^\dagger$ do not have $r-$dependence.  These operators are related to $\bm{\delta}^{(0)}$, $\bm{\Delta}^{(0)}$, and $\bm{D}^{(0)}$ via
\begin{align}
    &\bm{\delta}^{(0)} + 2n\beta^{(0)}  = \frac{1}{\sqrt{2}\Gamma}\mathcal{L}_n^\dagger, \quad \bm{\bar{\delta}}^{(0)} + 2n\bar{\beta}^{(0)}  = \frac{1}{\sqrt{2}\bar{\Gamma}}\mathcal{L}_n, \notag \\
    &
    \bm{\Delta}^{(0)}-2n(\gamma-\mu)^{(0)} = \bm{\Delta}^{(0)}-2n(\bar{\gamma}-\bar{\mu})^{(0)} = - \frac{\Delta}{2\Sigma} \mathcal{D}_n^\dagger, \notag \\
    & \bm{D}^{(0)}=\mathcal{D}_0. \label{eq:NR_der_op_Chan_op}
\end{align}
While acting on $e^{im\phi -i\omega t}$, they reduce to
\begin{align}
    &\mathcal{D}_n=\partial_r-\frac{iK}{\Delta}+2n\frac{r-1}{\Delta}, \quad \mathcal{D}_n^\dagger=\partial_r+\frac{iK}{\Delta}+2n\frac{r-1}{\Delta}, \notag \\
    &\mathcal{L}_n^\dagger=\partial_\theta-{P}+n\cot\theta, \notag 
\end{align}
with 
\begin{align}
    &K=(r^2+a^2)\omega-am,
    &{P}=-a\omega \sin\theta+m\csc\theta.
\end{align}
Note that here we adopt Teukolsky's convention for $K$ \cite{Teukolsky:1973ApJ}, which differs from Chandrasekhar's by a minus sign.

In our following calculations, we frequently  encounter two combinations of operators
\begin{align}
    &\bm{\delta}^{(0)}-p\beta^{(0)}-q\bar{\alpha}^{(0)}+m\bar{\pi}^{(0)}+n\tau^{(0)},\notag \\
    &\bm{\Delta}^{(0)}-p\gamma^{(0)}-q\bar{\gamma}^{(0)}+m\bar{\mu}^{(0)}+n\mu^{(0)}, \notag 
\end{align}
with $p,q,m,n$ being integers. They can be converted to
\begin{subequations}
\begin{align}
    &\frac{1}{\sqrt{2}\Gamma}\left[\mathcal{L}^\dagger_{(q-p)/2}-\left(\frac{m-q}{\Gamma}+\frac{n}{\bar{\Gamma}}\right)ia\sin\theta\right], \\
    &-\frac{\Delta}{2\Sigma}\left[\mathcal{D}^\dagger_{(p+q)/2}+\frac{(m-q)}{\Gamma}+\frac{(n-p)}{\bar{\Gamma}}\right],
\end{align}
\end{subequations}
respectively. In fact, the subscript $(p-q)/2$ and $(p+q)/2$ are called spin and boost weight in the Geroch-Held-Penrose (GHP) formalism \cite{penrose1984spinors}. Normally, these operators are applied sequentially. The factors  $\Delta,\Gamma,\Sigma,\sin\theta$ can be pulled out via the following commutation relations \cite{chandrasekhar1998mathematical}
\begin{equation}
\label{eq:commutation_relation_D_L}
\begin{aligned}
    &\Delta \mathcal{D}_{n+1}=\mathcal{D}_n \Delta, \quad \Delta \mathcal{D}_{n+1}^\dagger=\mathcal{D}_n^\dagger \Delta, \\
    &\left[\mathcal{D}_{m}^\dagger,\Gamma^n\right]=n \Gamma^{n-1}, \quad \left[\mathcal{D}_{m}^\dagger,\bar{\Gamma}^n\right]=n \bar{\Gamma}^{n-1},\\
    &\sin\theta\mathcal{L}_{n+1}=\mathcal{L}_{n}\sin\theta,  \quad \sin\theta\mathcal{L}^\dagger_{n+1}=\mathcal{L}^\dagger_{n}\sin\theta,  \\
    &\left[\mathcal{L}_{m}^\dagger,\Gamma^n\right]=-ina \Gamma^{n-1}\sin\theta, \quad \left[\mathcal{L}_{m}^\dagger,\bar{\Gamma}^n\right]=ina \bar{\Gamma}^{n-1}\sin\theta.
\end{aligned}
\end{equation}
Notably, the commutation does not change the structure 
\begin{align}
    \left[\mathcal{L}_{\ldots}^\dagger-\left(\frac{\ldots}{\Gamma}+\frac{\ldots}{\bar{\Gamma}}\right)ia\sin\theta\right], \quad     \left[\mathcal{D}_{\ldots}^\dagger+\left(\frac{\ldots}{\Gamma}+\frac{\ldots}{\bar{\Gamma}}\right)\right]. \notag 
\end{align}
Therefore, they can be used as (the only) two building blocks for $\theta$ and $r$ derivatives in the procedure of metric reconstruction. This is a major advantage of $\mathcal{D}_n$, $\mathcal{L}_n$ and their daggers over $\bm{\delta}^{(0)}$, $\bm{\Delta}^{(0)}$, and $\bm{D}^{(0)}$.

For completeness, we also provide the expressions of GHP operators \cite{penrose1984spinors}: 
\begin{equation}
\begin{aligned}
    &\text{\thorn}=\bm{D}^{(0)}-p\epsilon^{(0)}-q\bar{\epsilon}^{(0)},\\
    &\text{\thorn}^\prime=\bm{\Delta}^{(0)}-p\gamma^{(0)}-q\bar{\gamma}^{(0)}, \\
    &\eth_{\rm GHP}=\bm{\delta}^{(0)}-p\beta^{(0)}-q\bar{\alpha}^{(0)},\\
    &\eth^\prime_{\rm GHP}=\bm{\bar{\delta}}^{(0)}-p\alpha^{(0)}-q\bar{\beta}^{(0)},
\end{aligned} 
\end{equation}
in terms of Chandrasekhar's:
\begin{equation}
\begin{aligned}
    &\text{\thorn}=\mathcal{D}_0,\\
    &\text{\thorn}^\prime=-\frac{\Delta}{2\Sigma}\left(\mathcal{D}^\dagger_{(p+q)/2}-\frac{q}{\Gamma}-\frac{p}{\bar{\Gamma}}\right), \\
    &\eth_{\rm GHP}=\frac{1}{\sqrt{2}\Gamma}\left[\mathcal{L}^\dagger_{(q-p)/2}+\frac{iqa\sin\theta}{\Gamma}\right],\\
    &\eth^\prime_{\rm GHP}=\frac{1}{\sqrt{2}\bar{\Gamma}}\left[\mathcal{L}_{(p-q)/2}-\frac{ipa\sin\theta}{\bar{\Gamma}}\right].
\end{aligned} 
\end{equation}
Here the prime operation corresponds to the exchange of spinor basis \cite{penrose1984spinors}. 

Finally, for Schwarzschild BHs, $\mathcal{L}^\dagger_{s}$ and $\mathcal{L}_{s}$ become the $\eth$ operators for spin-weighted spherical harmonics
\begin{equation}
\label{eq:chan_L_vs_eth}
\begin{aligned}
    &-\mathcal{L}^\dagger_{-s}=\eth =-\left(\partial_\theta-m\csc\theta-s\cot\theta\right),  \\
    &-\mathcal{L}_s=\bar{\eth}=-\left(\partial_\theta+m\csc\theta+s\cot\theta\right).  
\end{aligned}
\end{equation}

\subsection{Teukolsky equation}
\label{subsec:Teukolsky_review}
To the linear order, the perturbations of $\Psi_{0,4}$ are described by the Teukolsky equation \cite{PhysRevLett.29.1114,Teukolsky:1973ApJ}
\begin{align}
    \mathcal{T} \left[\tensor[_{s}]{\psi}{^{(1)}}\right] = 0,
\end{align}
where $\tensor[_{2}]{\psi}{^{(1)}}=\Psi_0^{(1)}$ and $\tensor[_{-2}]{\psi}{^{(1)}}=\bar{\Gamma}^4\Psi_4^{(1)}$, meanwhile we
assume there are no matter sources. The Teukolsky operator $\mathcal{T}$ reads
\begin{widetext}
\begin{align}
\mathcal{T} =& \left [ \frac{(r^2+a^2)^2}{\Delta} -a^2\sin^2\theta\right ]\frac{\partial^2}{\partial t^2}+\frac{4 a r}{\Delta } \frac{\partial^2}{\partial t\partial \phi}+\left [\frac{a^2}{\Delta}-\frac{1}{\sin^2\theta}\right ]\frac{\partial^2}{\partial \phi^2} -\Delta^{-s}\frac{\partial}{\partial r} \left ( \Delta^{s+1}\frac{\partial}{\partial r}\right ) -\frac{1}{\sin\theta}\frac{\partial}{\partial \theta} \left ( \sin\theta \frac{\partial}{\partial \theta}\right ) \nonumber \\
&-2 s \left [ \frac{a(r-1)}{\Delta}+\frac{i \cos\theta}{\sin^2\theta}\right ]\frac{\partial}{\partial \phi}-2 s \left [\frac{r^2-a^2}{\Delta}-r-i a \cos\theta\right ]\frac{\partial}{\partial t}+(s^2\cot^2\theta-s). \label{eq:Teukolsky_operator_t_phi}
\end{align}
\end{widetext}
Decomposing $\tensor[_{s}]{\psi}{^{(1)}}$ into time and angular harmonics 
\begin{align}
    \tensor[_{s}]{\psi}{^{(1)}}= \tensor[_{s}]{{R_{lm\omega}^{(1)}}}{}(r) \tensor[_{s}]{{S_{lm\omega}}}{}(\theta) e^{i(m\phi-\omega t)},
\end{align}
the radial function $\tensor[_{s}]{{R_{lm\omega}^{(1)}}}{}(r)$ satisfies \cite{chandrasekhar1998mathematical}
\begin{subequations}
\label{eq:linear_teukolsky_s_pm2}
\begin{align}
    &( \Delta \mathcal{D}_1\mathcal{D}_2^\dagger+6i\omega r)\tensor[_{+2}]{{R_{lm\omega}^{(1)}}}{}=\tensor[_{+2}]{{\lambda_{lm\omega}^{(1)}}}{}\tensor[_{+2}]{{R_{lm\omega}^{(1)}}}{},  \label{eq:Teukolsky_equation_s2} \\
    &( \Delta \mathcal{D}_{-1}^\dagger\mathcal{D}_0-6i\omega r)\tensor[_{-2}]{{R_{lm\omega}^{(1)}}}{}=\tensor[_{-2}]{{\lambda_{lm\omega}^{(1)}}}{}\tensor[_{-2}]{{R_{lm\omega}^{(1)}}}{}. 
\end{align}
\end{subequations}
with $\tensor[_{\pm 2}]{{\lambda_{lm\omega}^{(1)}}}{}$ being the angular eigenvalues. By imposing the boundary conditions of a QNM, i.e., ingoing at the horizon $r_+$ and outgoing at infinity, an analytic representation for $\tensor[_{s}]{{R_{lm\omega}^{(1)}}}{}(r)$ can be constructed via Leaver's method \cite{leaver1985analytic}
\begin{align}
    \tensor[_{s}]{{R_{lm\omega}^{(1)}}}{} =& e^{i \omega r} (r-r_+)^{-s-i\sigma_+}(r-r_-)^{-1-s+2i \omega +i \sigma_+} \nonumber \\
    &\times \sum^\infty_{n=0} a_n \left( \frac{r-r_+}{r-r_-}\right)^n, \label{eq:leaver_method_s}
\end{align}
with $\sigma_+ =(\omega r_+-a m/2)/\sqrt{1-a^2}$. The coefficients $a_n$ are determined by three-term recurrence relations
\begin{align}
    & \alpha_0 a_1 +\beta_0 a_0 = 0, \quad \alpha_n a_{n+1}+\beta_n a_n+\gamma_n a_{n-1}=0. \label{eq:leaver_method_three_term_recurrence}
\end{align} 
where $\alpha_n$, $\beta_n$, and $\gamma_n$ are given in Appendix \ref{app:Leaver}. For our later convenience, we define 
\begin{align}
    &X_L(r)= e^{i \omega r} (r-r_+)^{-i\sigma_+}(r-r_-)^{2i \omega +i \sigma_+}\notag \\
    &=e^{i\omega \left(r_*-4r_+ u/a\right)+imu},  \label{eq:Leaver_XL}
\end{align}
with
\begin{align}
    \frac{dr_*}{dr}=\frac{r^2+a^2}{\Delta}, \quad \frac{du}{dr}=\frac{a}{\Delta}, \label{eq:def_rstar_u}
\end{align}
namely 
\begin{align}
    &r_*=r+\frac{r_+}{\sqrt{1-a^2}}\ln \left(\frac{r-r_+}{r_+}\right)-\frac{r_-}{\sqrt{1-a^2}}\ln \left(\frac{r-r_-}{r_-}\right), \notag \\
    &u=\frac{a}{2\sqrt{1-a^2}}\ln\frac{r-r_+}{r-r_-}. \notag 
\end{align}
The radial derivative of $X_L$ is given by
\begin{align}
    X_L^\prime (r)=\frac{iX_L(r)}{\Delta}\left[am+\omega (a^2+r^2-4r_+)\right].
\end{align}

\section{Reconstruction}
\label{sec:metric_reconstruction}

After specifying the Weyl scalar $\Psi_4^{(1)}$ for a QNM to the linear order, a crucial step toward second-order perturbations involves identifying a set of NP variables and a metric that is consistent with $\Psi_4^{(1)}$. This process is generally referred to as \textit{metric reconstruction}. The method was pioneered by Chrzanowski \cite{1975PhRvD..11.2042C} and Cohen and Kegeles \cite{COHEN19755,1979PhRvD..19.1641K} for source-free scenarios. In the presence of arbitrary matter sources, the method was extended recently by Green \etal \cite{Green:2019nam}. The CCK method centers around constructing a Hertz potential $\Psi$ as a solution to the Teukolsky equation. Subsequently, from $\Psi$, a perturbed metric and its corresponding spin coefficients can be derived, satisfying the linearized Einstein equations on top of a Kerr background. Wald's elegant argument \cite{PhysRevLett.41.203} attests to the viability of this methodology. However, a limitation of this approach lies in the challenge of directly inverting the differential equation to obtain the Hertz potential, given a particular $\Psi_4^{(1)}$. A workaround involves bypassing the use of the Hertz potential and performing reconstruction directly from $\Psi_4^{(1)}$ \cite{Loutrel:2020wbw, Ripley:2020xby}. This alternative method requires solving a set of differential equations.

In our specific scenario, namely the second-order effect of a QNM, the limitations associated with CCK are effectively eliminated. First, without the presence of matter sources, the complexity introduced by the stress tensor \cite{Green:2019nam} becomes irrelevant and the Hertz potential provides a comprehensive description for the perturbed metric (modulo gauge freedom and ``zero modes''). Second, as demonstrated below in Sec.~\ref{subsec:determining_hertz_potential}, the Hertz potential can be uniquely determined by $\Psi_4^{(1)}$ and the boundary condition of a QNM. Additionally, a notable advantage of CCK is its exclusive dependence on the differentiation of the Hertz potential, in contrast to the need for solving a set of partial derivative differential equations \cite{Loutrel:2020wbw, Ripley:2020xby}. This feature not only provides us with more analytical insights but also practical convenience in the numerical implementation. In fact, we will show in Sec.~\ref{subsec:metric_reconstruction} that the perturbed Weyl scalars, $\Psi_3^{(1)}$ and $\Psi_2^{(1)}$, can be succinctly expressed in terms of the Hertz potential.

The CCK procedure is accompanied by two gauge choices \cite{1975PhRvD..11.2042C}: the ingoing and outgoing radiation gauges. In particular, the outgoing radiation gauge (ORG) is asymptotically flat \cite{Campanelli:1998jv}, and its solution is directly related to physical observables, such as energy radiation and the excitation factor of second-order QNMs at future null infinity. Therefore, it is reasonable to adopt ORG to compute the gauge invariant excitation factor at infinity, even though the second-order perturbation $\Psi_4^{(2)}$ generally depends on the choice of gauge and tetrad \cite{Campanelli:1998jv}. 

Given the considerations above, we will adopt CCK and ORG for our subsequent calculations. Below, we first determine the Hertz potential for a linear QNM in Sec.~\ref{subsec:determining_hertz_potential}. Subsequently, in Sec.~\ref{subsec:metric_reconstruction}, we present the representation of the metric and Newman-Penrose variables in connection with the determined Hertz potential.

\subsection{Determining the Hertz potential}
\label{subsec:determining_hertz_potential}
In ORG, the Hertz potential $\hertz$ is related to  $\Psi_4^{(1)}$  via \cite{Whiting:2005hr,Keidl:2006wk,Keidl:2010pm} 
\begin{align}
    \Psi_4^{(1)}=\frac{1}{32}\frac{\Delta^4}{\bar{\Gamma}^4}\mathcal{D}_{2}^\dagger\mathcal{D}_{2}^\dagger\mathcal{D}_{2}^\dagger\mathcal{D}_2^\dagger\bar{\Psi}_{\rm ORG}, \label{eq:psi4_hertz}
\end{align}
where the bar stands for the complex conjugate. Since it is a linear differential equation of order four, $\bar{\Psi}_{\rm ORG}$ consists of a homogeneous part and a particular solution. First, the homogeneous part solves $\mathcal{D}_{2}^{\dagger\,4}\bar{\Psi}_{\rm hom}=0$, which is equivalent to 
\begin{align}
    \mathcal{D}_{0}^{\dagger\,4}\Delta^2\bar{\Psi}_{\rm hom}=0.
\end{align}
Here we have used the commutation relation $\Delta \mathcal{D}_{n+1}^\dagger=\mathcal{D}_n^\dagger \Delta$ in Eq.~\eqref{eq:commutation_relation_D_L}. 
Following \cite{Ori:2002uv}, a general solution reads 
\begin{align}
    \Delta^2\bar{\Psi}_{\rm hom}= \int d\omega e^{-i\omega (t+r_*)} \sum_m\sum_{i=0}^3 e^{im(\phi+u)} B_{im \omega}(\theta) r^i,
\end{align}
where $B_{im \omega}(\theta)$'s are arbitrary functions of $\theta$ and $u$ is defined in Eq.~\eqref{eq:def_rstar_u}. We can see that there are only ingoing modes depending on $t+r_*$. Since the Hertz potential is related to the Weyl scalar $\Psi_0^{(1)}$ through only time- and angular-derivatives  \cite{Whiting:2005hr,Keidl:2010pm}
\begin{align}
    \Psi_0^{(1)}=\frac{1}{8}\left[\mathcal{L}_{-1}^\dagger\mathcal{L}_{0}^\dagger\mathcal{L}_{1}^\dagger\mathcal{L}_{2}^\dagger \hertzbar+12\partial_t\hertz\right].
\end{align}
The homogenous part $\bar{\Psi}_{\rm hom}$ corresponds to a GW emerging from past null infinity, which should vanish in our pure-QNM context.

On the other hand, an efficient approach to determine the particular solution is to solve the $s=2$ source-free Teukolsky equation, which the Hertz potential $\hertz$ should satisfy in ORG \cite{1975PhRvD..11.2042C}, with the same QNM boundary conditions\footnote{Kerr BHs have the same QNM frequencies for $s=\pm 2$ perturbations.} for $\Psi_4^{(1)}$. Leveraging the Teukolsky-Starobinsky relation \cite{Starobinsky:1973aij,Teukolsky:1974yv}, we ensure that the solution consistently complies with Eq.~\eqref{eq:psi4_hertz}. As a result, the Hertz potential associated with $\Psi_4^{(1)}$ can be written as \cite{Berti:2005ys}
\begin{align}
    \hertz &= \sum_{lmn} B_{lm\omega}^{(a)}  \tensor[_{+2}]{{R_{lm\omega}}}{} (r)\, \tensor[_{+2}]{S}{_{lm\omega}}(\theta) e^{im\phi} e^{-i \omega_{lmn}t}  \notag \\
            & + B_{lm\omega}^{(b)} \tensor[_{+2}]{{\bar{R}}}{_{lm\omega}} (r) \tensor[_{+2}]{\bar{S}}{_{lm\omega}}(\pi-\theta) e^{-im\phi} e^{i \bar{\omega}_{lmn}t}, \label{eq:hertz_general_solution}
\end{align}
where $B_{lm\omega}^{(a)}$'s and $B_{lm\omega}^{(b)}$'s are constants, while $n$ stands for the radial overtone index. In the following discussions, we restrict ourselves to a single mode in $\Psi_4^{(1)}$, namely
\begin{align}
    \Psi_4^{(1)}\sim e^{i m \phi} e^{-i \omega_{lmn}(t-r_*)} \tensor[_{-2}]{S}{_{lm\omega}}(\theta), \quad r\to \infty,
\end{align}
which implies  
\begin{align}
    \hertzbar &= \tensor[_{+2}]{{R_{lm\omega}}}{} (r) \tensor[_{+2}]{S}{_{lm\omega}}(\pi-\theta) e^{im\phi} e^{-i \omega_{lmn}t} \notag \\
    &=(-1)^{m+l}\tensor[_{+2}]{{R_{lm\omega}}}{} (r) \tensor[_{-2}]{S}{_{lm\omega}}(\theta) e^{im\phi} e^{-i \omega_{lmn}t}, \label{eq:hertz_bar_qnm}
\end{align}
where we have used the property of spin-weighted spheroidal harmonics \cite{Nichols:2012jn}
\begin{align}
    \tensor[_{s}]{S}{_{lm\omega}}(\pi-\theta) = (-1)^{m+l}  \tensor[_{-s}]{S}{_{lm\omega}}(\theta).
\end{align}

As discussed in Sec.~\ref{subsec:Teukolsky_review}, Leaver's representation for $\tensor[_{+2}]{{R_{lm\omega}}}{}$ is given by
\begin{align}
    \tensor[_{+2}]{{R_{lm\omega}}}{} =X_L(r) (r-r_-)^{-1} \Delta^{-2} \sum^\infty_{n=0} a_n \left( \frac{r-r_+}{r-r_-}\right)^n, \label{eq:leaver_hertz}
\end{align}
At infinity, its asymptotic behavior reads
\begin{align}
    \tensor[_{+2}]{{R_{lm\omega}}}{} &=X_L(r) (r-r_-)^{-1} \Delta^{-2}\left(\sum^\infty_{n=0} a_n\right) \notag \\
    &\times\left[1+\frac{a_{\rm Hertz}}{r}+\mathcal{O}\left(\frac{1}{r^2}\right)\right],
\end{align}
with
\begin{align}
    &a_{\rm Hertz}=\frac{i}{2\omega_{lmn}} \tensor[_{+2}]{{\lambda_{lm\omega}^{(1)}}}{}-4+2iam-r_--4ir_+\omega_{lmn}.
\end{align}

\subsection{Determining the metric and Newman-Penrose variables}
\label{subsec:metric_reconstruction}
After obtaining the Hertz potential $\hertz$, we then follow CCK to compute the metric and NP variables. In particular, we will cast them into forms that are convenient for numerical calculations.

\subsubsection{Metric}
\label{subsubsec:reconstruction_metric}
In ORG, the nonvanishing metric components are $h_{\bar{m}\bar{m}}, h_{l\bar{m}}$, $h_{ll}$ and their complex conjugates. They are related to the Hertz potential through \cite{1975PhRvD..11.2042C}
\begin{subequations}
\label{eq:CCK_metric_original_expression}
\begin{align}
    &h_{ll}=\bar{\rho}^{-4}(\bm{\delta}-3\bar{\alpha}-\beta+5\bar{\pi})(\bm{\delta}-4\bar{\alpha}+\bar{\pi})\bar{\Psi}_{\rm ORG}+\rm{c}.\, c., \\
    &h_{\bar{m}\bar{m}}=\bar{\rho}^{-4}(\bm{\Delta}+5\bar{\mu}-3\bar{\gamma}+\gamma)(\bm{\Delta}+\bar{\mu}-4\bar{\gamma})\bar{\Psi}_{\rm ORG}, \\
    &h_{l\bar{m}}=\frac{1}{2}\bar{\rho}^{-4}(\bm{\delta}-3\bar{\alpha}+\beta+5\bar{\pi}+\tau)(\bm{\Delta}+\bar{\mu}-4\bar{\gamma}) \notag \\
    &+\frac{1}{2}\bar{\rho}^{-4}(\bm{\Delta}+5\bar{\mu}-\mu-3\bar{\gamma}-\gamma)(\bm{\delta}-4\bar{\alpha}+\bar{\pi})\bar{\Psi}_{\rm ORG}.
\end{align}
\end{subequations}
Here c.\,c. stands for the complex conjugate. With the help of Eq.~\eqref{eq:NR_der_op_Chan_op}, they become
\begin{subequations}
\label{eq:reconstruction_all_h}
\begin{align}
    &h_{ll}=\frac{\Gamma^2}{2}\left(\mathcal{L}_1^\dagger+\frac{2ia\sin\theta}{\Gamma}\right)\mathcal{L}_2^\dagger  \hertzbar+\rm{c}.\, c., \\
    &h_{\bar{m}\bar{m}}=\Delta^2\frac{\Gamma^2}{4\bar{\Gamma}^2}(\mathcal{D}_2^\dagger-\frac{2}{\Gamma})\mathcal{D}_2^\dagger\hertzbar, \label{eq:reconstruction_hmbarmbar} \\
    &h_{l\bar{m}}
    =-\frac{\Delta\Gamma^2}{2\sqrt{2}\bar{\Gamma}}\left[\frac{a^2\sin2\theta}{\Sigma}\mathcal{D}_2^\dagger+\left(\mathcal{D}_2^\dagger-\frac{2r}{\Sigma}\right)\mathcal{L}_2^\dagger\right]\bar{\Psi}_{\rm ORG}. \label{eq:reconstruction_hlmbar}
\end{align}
\end{subequations}
For a Schwarzschild BH, the expressions are reduced to
\begin{equation}
\begin{aligned}
    &h_{ll}=\frac{r^2}{2}\eth\eth \hertzbar+\rm{c}.\, c., \\
    &h_{\bar{m}\bar{m}}=\frac{\Delta^2}{4}\left(\mathcal{D}_2^\dagger-\frac{2}{r}\right)\mathcal{D}_2^\dagger\hertzbar, \\
    &h_{l\bar{m}}
    =\frac{r\Delta}{2\sqrt{2}}\left(\mathcal{D}_2^\dagger-\frac{2}{r}\right)\eth\bar{\Psi}_{\rm ORG},
\end{aligned}
\end{equation}
where we have used Eq.~\eqref{eq:chan_L_vs_eth}.

\subsubsection{Directional derivatives and spin coefficients}
\label{subsubsec:reconstruction_sc_dd}
In ORG, the first-order directional derivatives are given by \cite{Campanelli:1998jv}
\begin{subequations}
\begin{align}
    &\bm{D}^{(1)}=-\frac{1}{2}h_{ll}\bm{\Delta}^{(0)}, \\
    &\bm{\Delta}^{(1)}=0, \\
    &\bm{\delta}^{(1)}=-h_{lm}\bm{\Delta}^{(0)}+\frac{1}{2}h_{mm}\bm{\bar{\delta}}^{(0)}. \label{eq:directional_derivatives_delta1}
\end{align}
\end{subequations}
From Eq.~\eqref{eq:CCK_metric_original_expression} we can see that $\bm{\delta}^{(1)}$ is contributed exclusively by $\hertz$ and not by its complex conjugate. Similarly, the spin coefficients $\beta^{(1)}$, $\bar{\alpha}^{(1)}$, $\tau^{(1)}$, and $\bar{\pi}^{(1)}$ are given by \cite{Campanelli:1998jv}
\begin{subequations}
\begin{align}
    \beta^{(1)}=&-\frac{1}{4}(\bm{\Delta}+2\gamma+\mu+2\bar{\mu})^{(0)}h_{lm}\notag \\
    &+\frac{1}{4}(\bm{\bar{\delta}}+2\bar{\beta}-\pi-\bar{\tau})^{(0)}h_{mm}, \notag \\
    \bar{\alpha}^{(1)}=&-\frac{1}{4}(\bm{\bar{\delta}}-2\alpha+\pi+\bar{\tau})^{(0)}h_{mm}\notag \notag \\
    &-\frac{1}{4}(\bm{\Delta}+4\bar{\gamma}-2\gamma+\mu-2\bar{\mu})^{(0)}h_{lm}, \notag \\
    \tau^{(1)}=&\frac{1}{2}(\bm{\Delta}-2\gamma+\mu)^{(0)}h_{lm}-\frac{1}{2}\pi^{(0)} h_{mm},\notag \\
    \bar{\pi}^{(1)}=&-\frac{1}{2}(\bm{\Delta}-2\gamma+\mu)^{(0)}h_{lm}-\frac{1}{2}\bar{\tau}^{(0)} h_{mm}.\notag
\end{align}
\end{subequations}
They do not depend on $\hertzbar$ either. Here we emphasize that the expressions above have been simplified in ORG. See e.g., \cite{Loutrel:2020wbw} for the full expressions.

Meanwhile, three spin coefficients vanish identically in ORG  \cite{Campanelli:1998jv}
\begin{align}
    \nu^{(1)}=\gamma^{(1)}=\mu^{(1)}=0.
\end{align}
The expressions of other spin coefficients are not informative, therefore we include them in Appendix \ref{app:spin_coefficients}.

\subsubsection{\texorpdfstring{$\Psi_2$}{Psi2}}
$\Psi_2$ is intricately linked to the Hertz potential, e.g., see the ingoing-radiation-gauge counterpart in Eq.~(97) of Ref.~\cite{Keidl:2006wk}. Nevertheless, we find that its expression can be significantly simplified to
\begin{align}
&\frac{16\bar{\Gamma}^4}{\Delta^2}\Psi_2^{(1)}= \bar{\Gamma}^2\left(\mathcal{D}^\dagger_2-\frac{2}{\bar{\Gamma}}\right)^2 \mathcal{L}_1^\dagger\mathcal{L}_2^\dagger\hertzbar \notag \\
&-4ia\sin\theta \bar{\Gamma} \left(\mathcal{D}^\dagger_2-\frac{2}{\bar{\Gamma}}\right) \left(\mathcal{D}^\dagger_2-\frac{1}{\bar{\Gamma}}\right) \mathcal{L}_2^\dagger\hertzbar \notag \\
&-6a^2\sin^2\theta \left(\mathcal{D}^\dagger_2-\frac{2}{\bar{\Gamma}}\right)\mathcal{D}^\dagger_2\hertzbar. \label{eq:psi2_hertz_final}
\end{align}
In particular, for Schwarzschild BHs, we have
\begin{align}
&\frac{16r^2}{\Delta^2}\Psi_2^{(1)}=\left(\mathcal{D}^\dagger_2-\frac{2}{r}\right)^2 \eth\eth\hertzbar.
\end{align}
Details of the derivation can be found in Appendix \ref{app:psi2_hertz}.


\subsubsection{\texorpdfstring{$\Psi_3$}{Psi3}}
To obtain $\Psi_3$, we start from the Ricci identity \cite{chandrasekhar1998mathematical}
\begin{align}
    \Psi_3=(\bm{\bar{\delta}}+\bar{\beta}-\bar{\tau})\gamma-(\bm{\Delta}-\bar{\gamma}+\bar{\mu})\alpha+(\rho+\epsilon)\nu-(\tau+\beta)\lambda.
\end{align}
Linearizing it leads to
\begin{align}
    \Psi_3^{(1)}&=(\bm{\bar{\delta}}+\bar{\beta}-\bar{\tau})^{(1)}\gamma^{(0)}-(\bm{\Delta}-\bar{\gamma}+\bar{\mu})^{(0)}\alpha^{(1)} \notag \\
    &-(\tau+\beta)^{(0)}\lambda^{(1)},
\end{align}
since  $\mu^{(1)}=\gamma^{(1)}=\nu^{(1)}=\bm{\Delta}^{(1)}=\nu^{(0)}=\lambda^{(0)}=0$. We can then express $\Psi_3^{(1)}$ in terms of the Hertz potential by using the results given in Secs.~\ref{subsubsec:reconstruction_metric} and \ref{subsubsec:reconstruction_sc_dd}. This yields a tedious expression, e.g., see the ingoing-radiation-gauge counterpart in Eq.~(98) of Ref.~\cite{Keidl:2006wk}. However, by virtue of Chandrasekhar's operators $\mathcal{D}_n^\dagger$ and $\mathcal{L}_n^\dagger$, and the commutator in Eq.~\eqref{eq:chandra_D_commutator}, we find the expression can be simplified straightforwardly into
\begin{align}
 &\frac{16\sqrt{2}\bar{\Gamma}^3}{\Delta^3}\Psi_3^{(1)} 
=-\left(\mathcal{L}_2^\dagger-\frac{3ia\sin\theta}{\bar{\Gamma}}\right)\left(\mathcal{D}^\dagger_2-\frac{1}{\bar{\Gamma}} \right)^3\,\hertzbar \notag \\
&+\frac{6ia\sin\theta}{\bar{\Gamma}^2}\left(\mathcal{D}^\dagger_2-\frac{2}{\bar{\Gamma}} \right)^2\,\hertzbar.
\end{align}
For Schwarzschild BHs, it reduces to
\begin{align}
 &\frac{16\sqrt{2}r^3}{\Delta^3}\Psi_3^{(1)} 
=\left(\mathcal{D}^\dagger_2-\frac{1}{r} \right)^3\,\eth\hertzbar.
\end{align}

\subsubsection{\texorpdfstring{$\Psi_4$}{Psi4}}
Finally, to compute $\Psi_4$, we adopt the Ricci identity \cite{chandrasekhar1998mathematical}
\begin{align}
    \Psi_4=(\bm{\bar{\delta}}+3\alpha+\bar{\beta}+\pi-\bar{\tau})\nu-(\bm{\Delta}+\mu+\bar{\mu}+3\gamma-\bar{\gamma})\lambda.
\end{align}
By using $\nu^{(0)}=\lambda^{(0)}=\nu^{(1)}=0$ and $\mu^{(0)}-\gamma^{(0)}=\bar{\mu}^{(0)}-\bar{\gamma}^{(0)}$,
the expression becomes
\begin{align}
    \Psi_4^{(1)}=-(\bm{\Delta}+2\mu+2\gamma)^{(0)}\lambda^{(1)}.
\end{align}
We then insert the expression of $\lambda^{(1)}$ in  Eq.~\eqref{eq:lambda_1st_order} to reproduce the known result in the literature \cite{Whiting:2005hr,Keidl:2006wk,Keidl:2010pm}
\begin{align}
    \Psi_4^{(1)}=\frac{1}{32}\frac{\Delta^4}{\bar{\Gamma}^4}\mathcal{D}_{2}^\dagger\mathcal{D}_{2}^\dagger\mathcal{D}_{2}^\dagger\mathcal{D}_2^\dagger\bar{\Psi}_{\rm ORG}. \label{eq:reconstruction_psi4}
\end{align}
This serves as a consistency check for our previous steps to determine the Hertz potential, metric, and spin coefficients.
After applying Leaver's representation for $\hertzbar$ in Eq.~\eqref{eq:leaver_hertz}, we obtain the asymptotic expression of the 
$(l,m)$ harmonic of $\Psi_4^{(1)}$ at infinity 
\begin{align}
    \left[r\Psi_4^{(1)}\right]_{l,m}^{(\infty)}=\frac{\omega_{lmn}^4}{2} e^{-i\omega_{lmn} (t-r_*)} \sum_{n=0}^\infty a_n. \label{eq:linear_psi4_inf}
\end{align}

\section{The second-order Teukolsky equation}
\label{sec:second_order_Teukolsky_equation}
It is known that the second order perturbation $\Psi_4^{(2)}$ still follows the Teukolsky equation \cite{Campanelli:1998jv}
\begin{align}
    \mathcal{T} \left[\Psi_4^{(2)}\right] = S_4^{(2)}, \label{eq:2nd_Teukolsky}
\end{align}
where the source term $S_4^{(2)}$ is given by
\begin{align}
    S_4^{(2)}=&-\left[d_4^{(0)}(\bm{D}+4\epsilon-\rho)^{(1)}-d_3^{(0)}(\bm{\delta}+4\beta-\tau)^{(1)}\right] \Psi_4^{(1)} \notag \\
    &+ \left[d_4^{(0)}(\bm{\bar{\delta}}+2\alpha+4\pi)^{(1)}-d_3^{(0)}(\bm{\Delta}+2\gamma+4\mu)^{(1)}\right] \Psi_3^{(1)} \notag \\
    & -3\left[d_4^{(0)}\lambda^{(1)}-d_3^{(0)}\nu^{(1)}\right]\Psi_2^{(1)}\notag \\
    &-3\Psi_2^{(0)}\left[\left(d_4-3\mu\right)^{(1)}\lambda^{(1)}-\left(d_3-3\pi\right)^{(1)}\nu^{(1)}\right],
\end{align} 
and the two operators $d_{3,4}$ read
\begin{subequations}
\begin{align}
    &d_3=\bm{\bar{\delta}}+3\alpha+\bar{\beta}+4\pi-\bar{\tau}, \\ 
    &d_4=\bm{\Delta}+4\mu+\bar{\mu}+3\gamma-\bar{\gamma}.
\end{align} 
\end{subequations}
From Sec.~\ref{subsubsec:reconstruction_sc_dd}, we notice that $\nu^{(1)}=\mu^{(1)}=\gamma^{(1)}=\bm{\Delta}^{(1)}=0$ in ORG, which yields $d_4^{(1)}=0$. As a result, the source reduces to
\begin{align}
    S_4^{(2)}=&-\left[d_4^{(0)}(\bm{D}+4\epsilon-\rho)^{(1)}-d_3^{(0)}(\bm{\delta}+4\beta-\tau)^{(1)}\right] \Psi_4^{(1)} \notag \\
    &+ d_4^{(0)}\left[(\bm{\bar{\delta}}+2\alpha+4\pi)^{(1)}\Psi_3^{(1)}-3\lambda^{(1)}\Psi_2^{(1)}\right]. \notag 
\end{align}

Since all the reconstructed quantities depend linearly on $\hertz$ and its complex conjugate $\hertzbar$, they lead to three types of quadratic effects:
\begin{align}
    \hertz^2\sim e^{2i\bar{\omega}_{lmn}t},\quad \hertz\hertzbar\sim e^{i(\bar{\omega}_{lmn}-\omega_{lmn})t},
\end{align}
and
\begin{align}
    \hertzbar^2 \sim e^{-2i\omega_{lmn}t}. \label{eq:quadratic_prograde}
\end{align}
Our subsequent calculations focus exclusively on the excitation of the mode $\sim e^{-2i\omega_{lmn}t}$ in Eq.~\eqref{eq:quadratic_prograde}. Therefore,
terms that depend on $\hertz$ do not contribute to the final second-order Teukolsky equation. Accordingly, in the expressions below, we will simply drop terms of $\hertz$. Whenever this is done we will
indicate that such terms have been dropped by using the symbol $\approx$ instead of =. For instance, in Sec.~\ref{subsubsec:reconstruction_sc_dd}, we notice that $\bm{\delta}^{(1)}$ is fully contributed by $\hertz$, the term $\bm{\delta}^{(1)}\Psi_4^{(1)}$ in $S_4^{(2)}$ is irrelevant to the excitation of $\sim e^{-2i\omega_{lmn}t}$, namely $\bm{\delta}^{(1)}\Psi_4^{(1)}\approx 0$. Similarly, we have $\beta^{(1)} \approx\tau^{(1)}\approx0$. With this observation, the source $S_4^{(2)}$ further reduces to
\begin{align}
    S_4^{(2)}\approx &d_4^{(0)}\left[-(\bm{D}+4\epsilon-\rho)^{(1)} \Psi_4^{(1)} + (\bm{\bar{\delta}}+2\alpha+4\pi)^{(1)} \Psi_3^{(1)}\right. \notag \\
    &\left. -3\lambda^{(1)}\Psi_2^{(1)}\right]. \label{eq:final_source_from_NR_variables}
\end{align}
At this stage, we can express $S_4^{(2)}$ in terms of the Hertz potential by inserting the reconstructed variables in Sec.~\ref{sec:metric_reconstruction}, and obtain
\begin{widetext}
\begin{align}
    &2\bar{\Gamma}^4\Sigma S_4^{(2)} \times\frac{16}{\Delta^6}\approx  e^{-2i\omega_{lmn}t+2im\phi}\left\{ A_1 \left[\mathcal{L}_2^\dagger S(\theta)\right]^2 + A_2 \left[S(\theta)\,\mathcal{L}_1^\dagger\mathcal{L}_2^\dagger S(\theta)\right]+ A_3 \left[3ia\sin\theta\,S \, \mathcal{L}_2^\dagger S(\theta) \right] + A_4 \left[a\sin\theta S(\theta)\right]^2\right\}, \label{eq:final_source_term}
\end{align}
with
\begin{align}
A_1=& -\frac{45 R^2 \Gamma}{2 \bar{\Gamma}^7} + \frac{3(15 \Gamma-\bar{\Gamma}) R \mathcal{D}^{\dagger}_{2}R}{\bar{\Gamma}^6} + \frac{3(\bar{\Gamma}-15\Gamma) \left(\mathcal{D}^{\dagger}_{2}R\right)^2}{2\bar{\Gamma}^5}  + \frac{(3\bar{\Gamma}-42\Gamma) R \mathcal{D}^{\dagger\,2}_{2}R}{2 \bar{\Gamma}^5}  + \frac{6 R \Gamma \mathcal{D}^{\dagger\,3}_{2}R}{\bar{\Gamma}^4} + \frac{3(7 \Gamma-\bar{\Gamma}) \mathcal{D}^{\dagger}_{2}R \mathcal{D}^{\dagger\,2}_{2}R}{\bar{\Gamma}^4}  \notag \\
&+ \frac{3(2\bar{\Gamma}-13\Gamma) \left(\mathcal{D}^{\dagger\,2}_{2}R\right)^2}{8 \bar{\Gamma}^3}  + \frac{(\bar{\Gamma}-12\Gamma)\mathcal{D}^{\dagger}_{2}R \mathcal{D}^{\dagger\,3}_{2}R}{2 \bar{\Gamma}^3} - \frac{(7 \Gamma +4r) R \mathcal{D}^{\dagger\,4}_{2}R}{8 \bar{\Gamma}^3}     + \frac{(11 \Gamma-\bar{\Gamma}) \mathcal{D}^{\dagger\,2}_{2}R \mathcal{D}^{\dagger\,3}_{2}R}{4 \bar{\Gamma}^2}   + \frac{(4 \Gamma+r) \mathcal{D}^{\dagger}_{2}R \mathcal{D}^{\dagger\,4}_{2}R}{4 \bar{\Gamma}^2}\notag \\
&+ \frac{rR \mathcal{D}^{\dagger\,5}_{2}R}{4 \bar{\Gamma}^2}   - \frac{3\Gamma\left( \mathcal{D}^{\dagger\,3}_{2}R\right)^2}{8 \bar{\Gamma}}  - \frac{\Gamma \mathcal{D}^{\dagger\,2}_{2}R \mathcal{D}^{\dagger\,4}_{2}R}{2 \bar{\Gamma}}  - \frac{\Gamma \mathcal{D}^{\dagger}_{2}R \mathcal{D}^{\dagger\,5}_{2}R}{8 \bar{\Gamma}},
\end{align}

\begin{align}
    A_2 = &-\frac{3 R \mathcal{D}^{\dagger}_2 R}{\bar{\Gamma}^5} + \frac{3 r R  \mathcal{D}^{\dagger\,2}_2 R}{\bar{\Gamma}^5} + \frac{3 \left(\mathcal{D}^{\dagger}_2 R\right)^2}{\bar{\Gamma}^4}+ \frac{3 r \left(\mathcal{D}^{\dagger\,2}_2 R\right)^2 }{2 \bar{\Gamma}^3}  - \frac{3(2 r+ \bar{\Gamma}) \mathcal{D}^{\dagger}_2 R \mathcal{D}^{\dagger\,2}_2 R}{2\bar{\Gamma}^4}- \frac{3 R \Gamma \mathcal{D}^{\dagger\,3}_2 R}{2 \bar{\Gamma}^4} + \frac{(r+\Gamma) \mathcal{D}^{\dagger}_2 R \mathcal{D}^{\dagger\,3}_2 R}{\bar{\Gamma}^3} \notag \\
    &+ \frac{(3 \Gamma-\bar{\Gamma}) R  \mathcal{D}^{\dagger\,4}_2 R}{4 \bar{\Gamma}^3} - \frac{(2r+3\Gamma) \mathcal{D}^{\dagger\,2}_2 R \mathcal{D}^{\dagger\,3}_2 R}{4 \bar{\Gamma}^2} + \frac{(\bar{\Gamma}-6\Gamma)\mathcal{D}^{\dagger}_2 R  \mathcal{D}^{\dagger\,4}_2 R}{8 \bar{\Gamma}^2}+ \frac{(\bar{\Gamma}-2\Gamma) R  \mathcal{D}^{\dagger\,5}_2 R}{8 \bar{\Gamma}^2}+ \frac{\Gamma \left( \mathcal{D}^{\dagger\,3}_2 R\right)^2}{4 \bar{\Gamma}} + \frac{7 \Gamma \mathcal{D}^{\dagger\,2}_2 R \mathcal{D}^{\dagger\,4}_2 R}{16 \bar{\Gamma}} \notag \\
    & + \frac{\Gamma \mathcal{D}^{\dagger}_2 R \mathcal{D}^{\dagger\,5}_2 R}{4 \bar{\Gamma}} + \frac{R \Gamma \mathcal{D}^{\dagger\,6}_2 R}{16 \bar{\Gamma}},
\end{align}

\begin{align}
    &A_3=\frac{5(2\bar{\Gamma}-3 \Gamma) R \mathcal{D}^{\dagger}_2R}{\bar{\Gamma}^7} +
\frac{(15 \Gamma-11\bar{\Gamma}) \left(\mathcal{D}^{\dagger}_2R\right)^2}{\bar{\Gamma}^6} +
\frac{5(2 \Gamma- \bar{\Gamma}) R \mathcal{D}^{\dagger\,2}_2R}{\bar{\Gamma}^6}  +
\frac{(23\bar{\Gamma}-34\Gamma) \mathcal{D}^{\dagger}_2R \mathcal{D}^{\dagger\,2}_2R}{2\bar{\Gamma}^5}  +
\frac{(\bar{\Gamma}-6\Gamma)R \mathcal{D}^{\dagger\,3}_2R}{2\bar{\Gamma}^5} \notag \\
&+
\frac{3(3 \Gamma-2\bar{\Gamma})\left( \mathcal{D}^{\dagger\,2}_2R\right)^2}{2\bar{\Gamma}^4} +
\frac{(15 \Gamma-8\bar{\Gamma}) \mathcal{D}^{\dagger}_2R \mathcal{D}^{\dagger\,3}_2R}{3\bar{\Gamma}^4} +
\frac{(2 r+\Gamma) R \mathcal{D}^{\dagger\,4}_2R}{6\bar{\Gamma}^4}  +
\frac{(16\bar{\Gamma}-29\Gamma) \mathcal{D}^{\dagger\,2}_2R \mathcal{D}^{\dagger\,3}_2R}{12\bar{\Gamma}^3}  +
\frac{(4\bar{\Gamma}-21\Gamma)\mathcal{D}^{\dagger}_2R \mathcal{D}^{\dagger\,4}_2R}{24\bar{\Gamma}^3}\notag \\
&-
\frac{(3\bar{\Gamma}+2r) R \mathcal{D}^{\dagger\,5}_2R}{24\bar{\Gamma}^3}  +
\frac{(3\Gamma-4\bar{\Gamma})\left( \mathcal{D}^{\dagger\,3}_2R\right)^2}{12\bar{\Gamma}^2}  +
\frac{(8\Gamma-\bar{\Gamma}) \mathcal{D}^{\dagger\,2}_2R \mathcal{D}^{\dagger\,4}_2R}{24\bar{\Gamma}^2}  +
\frac{r \mathcal{D}^{\dagger}_2R \mathcal{D}^{\dagger\,5}_2R}{6\bar{\Gamma}^2} +
\frac{R \mathcal{D}^{\dagger\,6}_2R}{24\bar{\Gamma}},
\end{align}

\begin{align}
    &A_4=\frac{(45 \Gamma-60\bar{\Gamma}) \left(\mathcal{D}^{\dagger}_2R\right)^2}{2\bar{\Gamma}^7} +
\frac{(42\bar{\Gamma}-30 \Gamma) \mathcal{D}^{\dagger}_2R \mathcal{D}^{\dagger\,2}_2R}{\bar{\Gamma}^6}  +
\frac{9(2 \Gamma-3\bar{\Gamma}) \left(\mathcal{D}^{\dagger\,2}_2R\right)^2}{2\bar{\Gamma}^5} +
\frac{9(2 \Gamma-3\bar{\Gamma}) \mathcal{D}^{\dagger}_2R \mathcal{D}^{\dagger\,3}_2R}{2\bar{\Gamma}^5} \notag \\
& +
\frac{3(5\bar{\Gamma}-3\Gamma) \mathcal{D}^{\dagger\,2}_2R \mathcal{D}^{\dagger\,3}_2R}{2\bar{\Gamma}^4}+
\frac{(5\bar{\Gamma}-3\Gamma) \mathcal{D}^{\dagger}_2R \mathcal{D}^{\dagger\,4}_2R}{2\bar{\Gamma}^4}   +
\frac{3( \Gamma-2 \bar{\Gamma}) \left(\mathcal{D}^{\dagger\,3}_2R\right)^2}{8\bar{\Gamma}^3}  +
\frac{(\Gamma-2\bar{\Gamma}) \mathcal{D}^{\dagger\,2}_2R \mathcal{D}^{\dagger\,4}_2R}{2\bar{\Gamma}^3} +
\frac{(\Gamma-2\bar{\Gamma}) \mathcal{D}^{\dagger}_2R \mathcal{D}^{\dagger\,5}_2R}{8\bar{\Gamma}^3}.
\end{align}
Here $R$ refers to the radial function of the Hertz potential $\tensor[_{+2}]{{R_{lm\omega}}}{}$ in Eq.~\eqref{eq:leaver_hertz}, and $S(\theta)$ stands for $\tensor[_{-2}]{S}{_{lm\omega}}(\theta)$ in Eq.~\eqref{eq:hertz_bar_qnm}.
\end{widetext}
It is evident that $S_4^{(2)}$ cannot be decomposed into a structure of $(\rm{angular})\times(\rm{radial})$. Nevertheless, in Eq.~\eqref{eq:final_source_term}, we can still classify the terms based on four primary angular dependencies:
\begin{align}
    &\left[\mathcal{L}_2^\dagger S(\theta)\right]^2,   &\left[S(\theta)\,\mathcal{L}_1^\dagger\mathcal{L}_2^\dagger S(\theta)\right],  \notag \\
    &\left[a\sin\theta\,S(\theta) \mathcal{L}_2^\dagger S(\theta) \right], &\left[a\sin\theta S(\theta)\right]^2. \notag 
\end{align}
The angular dependence of their coefficients $A_{1..4}$ is solely determined by the scalar function $\Gamma=r+ia\cos\theta$ and its complex conjugate $\bar{\Gamma}$. In particular, we have
\begin{align}
    A_{1..4}\sim \frac{\Gamma^j}{\bar{\Gamma}^k},
\end{align}
with $j=0,1$ and $k=1..7$.

Since $S_4^{(2)}$ is non-separable, a linear QNM can induce quadratic effects in various spin-weighted spheroidal harmonics. To investigate a specific excitation channel, such as 
\begin{align}
    (l_L,m_L,n_L)\times (l_L,m_L,n_L) \to (l_Q,m_Q), \label{eq:excitation_channel}
\end{align}
angular projection of the second-order Teukolsky equation is necessary. In the rest of this section, we will accomplish this goal via two steps. First in Sec.~\ref{subsec:separation_Teukolsky}, we follow Teukolsky \cite{Teukolsky:1973ApJ} to decompose the Teukolsky operator on the left-hand side into angular and radial parts. Next in Sec.~\ref{subsec:angular_projection}, we use the property of spin-weighted spheroidal harmonics to perform the projection. Finally, in Secs.~\ref{subsec:angular_projection_Schwarzschild_BH} and \ref{subsec:angular_projection_Kerr_BH}, we explore the features of the angular projection for Schwarzschild and Kerr BHs, respectively. For conciseness, we will use the notation $\omega_L$ to stand for the complex frequency of the linear mode $\omega_{lmn}$.

\subsection{Separation of the Teukolsky operator}
\label{subsec:separation_Teukolsky}
We follow Teukolsky \cite{Teukolsky:1973ApJ} and rescale both sides of the equation by a factor of $2\bar{\Gamma}^4\Sigma$. This converts the variable to $\psi^{(2)}=\bar{\Gamma}^4\Psi_4^{(2)}$. Then we can write 
\begin{align}
    \psi^{(2)} = e^{-2i\omega_{L} t}e^{2im_L \phi}\sum_{l} R^{(2)}_{l}(r) \, S^{(2)}_{l}(\theta) . \label{eq:psi2_separation_of_the_Teukolsky_operator}
\end{align}
Notice that the time dependence has been set to $e^{-2i\omega_{L} t}$, to match that of the source term $\hertzbar^2\sim e^{-2i\omega_{L} t}$. The azimuthal quantum number has also been chosen to be twice the linear one $m_L$. In other words, Eq.~\eqref{eq:psi2_separation_of_the_Teukolsky_operator} represents a particular solution to the second-order perturbation equation. The homogenous part traces the evolution of linear QNMs, which is not pertinent to our studies. Plugging Eq.~\eqref{eq:psi2_separation_of_the_Teukolsky_operator} into Eq.~\eqref{eq:2nd_Teukolsky} yields
\begin{align}
    & e^{-2i\omega_L t+2im_L\phi}\sum_{l} \left[ S_{l}^{(2)}(\theta) (\Delta \mathcal{D}_{-1}^{\dagger\,(2)}\mathcal{D}^{(2)}_0-12i\omega_L r) R_{l}^{(2)}(r) \right. \notag \\
    &+\left.R_{l}^{(2)}(r)  (\mathcal{L}^{(2)}_{-1}\mathcal{L}_2^{\dagger\,(2)}+12a\omega_L \cos\theta) S_{l}^{(2)}(\theta) \right]  \notag \\
    & = - 2\bar{\Gamma}^4\Sigma S_4^{(2)} . \label{eq:2nd_Teukolsky_before_separation}
\end{align}
We note that the operators $\mathcal{D}^{(2)}$'s and $\mathcal{L}^{(2)}$'s (including their daggers) correspond to the second-order mode with the frequency $2\omega_L$. The angular basis $S_{l}^{(2)}$ can be chosen to satisfy 
\begin{align}
(\mathcal{L}^{(2)}_{-1}\mathcal{L}_2^{\dagger\,(2)}+12a\omega_L \cos\theta )S_{l}^{(2)} = -\lambda^{(2)}_l  S_{l}^{(2)}.
\end{align}
where $\lambda^{(2)}_l$ is the corresponding eigenvalue.
Then Eq.~\eqref{eq:2nd_Teukolsky_before_separation} reduces to 
\begin{align}
    & e^{-2i\omega_L t+2im\phi} \sum_{l} S_{l}^{(2)}(\theta) \notag \\
    &\times(\Delta \mathcal{D}_{-1}^{\dagger\,(2)}\mathcal{D}^{(2)}_0-12i\omega_L r-\lambda^{(2)}_l) R_{l}^{(2)}(r)  = - 2\bar{\Gamma}^4\Sigma S_4^{(2)} . \label{eq:2nd_Teukolsky_after_angular}
\end{align}
\subsection{Angular projection}
\label{subsec:angular_projection}

For a certain $\omega$ and $m$, the spin-weighted spheroidal harmonics associated with different $l$'s are orthogonal. To see this, we consider two different spin-weighted spheroidal harmonics $S_1$ and $S_2$, satisfying 
\begin{subequations}
\begin{align}
    &(\mathcal{L}_{-1}\mathcal{L}_2^\dagger+12a\omega_L \cos\theta )S_1 = \lambda_1  S_1,  \\
    &(\mathcal{L}_{-1}\mathcal{L}_2^\dagger+12a\omega_L \cos\theta )S_2 = \lambda_2  S_2.
\end{align}  
\end{subequations}
with $\lambda_1\neq \lambda_2$.
The equations above yield
\begin{align}
    S_2\, \mathcal{L}_{-1}\mathcal{L}_2^\dagger S_1- S_1\, \mathcal{L}_{-1}\mathcal{L}_2^\dagger S_2 = (\lambda_1-\lambda_2) S_1S_2.
\end{align}
Integrating both sides with the measure $\sin \theta$ leads to
\begin{align}
    &(\lambda_1-\lambda_2) \int_0^\pi  S_1S_2 \sin\theta d\theta \notag \\
    &= \int_0^\pi \left(S_2\, \mathcal{L}_{-1}\mathcal{L}_2^\dagger S_1- S_1\, \mathcal{L}_{-1}\mathcal{L}_2^\dagger S_2\right) \sin\theta d\theta.
\end{align}
The value of the right-hand-side vanishes due to a property of the angular operator $\mathcal{L}$ (see LEMMA 4 in Chapter 8 of \cite{chandrasekhar1998mathematical})
\begin{align}
    \int_0^\pi g(\mathcal{L}_nf)\sin\theta d\theta  = - \int_0^\pi f(\mathcal{L}_{1-n}^\dagger g) \sin\theta d\theta.
\end{align}
Therefore, we have 
\begin{align}
     \int_0^\pi  S_1S_2 \sin\theta d\theta = 0,
\end{align}
since $\lambda_1\neq \lambda_2$.

Consequently, to consider the excitation channel in Eq.~\eqref{eq:excitation_channel}, we can project both sides of Eq.~\eqref{eq:2nd_Teukolsky_after_angular} to the  angular basis of the child mode $S_{l_Q}(\theta)$ via:
\begin{align}
    & (\Delta \mathcal{D}_{-1}^{\dagger\,(2)}\mathcal{D}^{(2)}_0-12i\omega_L r-\lambda^{(2)}_{l_Q}) R_{l_Q}^{(2)}(r) \notag \\
    &=-e^{-im_Q\phi + 2i\omega_L t} \frac{\int_0^\pi2\bar{\Gamma}^4\Sigma S_4^{(2)} \times S_{l_{Q}}(\theta) \sin\theta d\theta}{\int_0^\pi S^2_{l_{Q}}(\theta) \sin\theta d\theta} \notag \\
    &\equiv Q(r), \label{eq:2nd_equation_with_angular_projection}
\end{align}
where we have defined the right-hand-side source to be $Q(r)$, which depends only on $r$. In Appendix \ref{app:source_Schwarzschild_l2}, we provide an explicit expression of $Q(r)$ for Schwarzschild BHs with $l_L=2$ and $l_Q=4$. By virtue of Leaver's representation in Eq.~\eqref{eq:leaver_hertz}, $Q(r)$ has an asymptotic expansion as follows 
\begin{align}
    Q(r)=&4r^2\omega_L^6X^2_L(r)\left(\sum^\infty_{n=0} a_n\right)^2 \left[1+\mathcal{O}\left(\frac{1}{r}\right)\right] \notag \\
    &\times\left<S_{l_L}\,\mathcal{L}_1^\dagger\mathcal{L}_2^\dagger S_{l_L}-\left(\mathcal{L}_2^\dagger S_{l_L}\right)^2\right|\left.S_{l_{Q}}\right>. \label{eq:Q_asymptotic_expansion}
\end{align}
where the angular inner product $\left<f\right|\left.S_{l_{Q}}\right>$ is defined to be
\begin{align}
    &\frac{\int_{0}^{ \pi} f\times S_{l_{Q}} \sin\theta d\theta}{\int_0^\pi S^2_{l_{Q}} \sin\theta d\theta}.  \label{eq:angular_inner_product}
\end{align}
We can see $Q(r)\sim \mathcal{O}(r^{2})$, an order lower than the left-hand-side of Eq.~\eqref{eq:2nd_equation_with_angular_projection}, which is $\mathcal{O}(r^3)$. With this at hand, we can solve Eq.~\eqref{eq:2nd_equation_with_angular_projection} order by order in the large$-r$ limit, which yields
\begin{align}
    R_{l_Q}^{(2)}\propto (r-r_+)X_L^2\Delta \left[1+\frac{a_Q}{r}+\mathcal{O}\left(\frac{1}{r^2}\right)\right]. \label{eq:RQ_asymptotic}
\end{align}
with
\begin{align}
    a_Q=2+4iam_L+r_++i\frac{\lambda^{(2)}_{l_Q}}{2\omega_Q}-8ir_+\omega_L.
\end{align}
The undetermined coefficient in Eq.~\eqref{eq:RQ_asymptotic} corresponds to the amplitude of the quadratic QNM, which is the target of this work.

\subsection{Schwarzschild black holes}
\label{subsec:angular_projection_Schwarzschild_BH}
For Schwarzschild BHs, the angular dependence of the source $S_4^{(2)}$ in Eq.~\eqref{eq:final_source_term} can be simplified significantly. In particular, by the virtue of 
\begin{subequations}
\begin{align}
    &\mathcal{L}_2^\dagger S(\theta)= -\eth\,  \tensor[_{-2}]{Y}{_{l_Lm_L}}=-\sqrt{(l_L+2)(l_L-1)}\, \tensor[_{-1}]{Y}{_{l_Lm_L}},\notag \\
    &\mathcal{L}_1^\dagger\mathcal{L}_2^\dagger S(\theta)=\eth\eth \tensor[_{-2}]{Y}{_{l_Lm_L}} \notag \\
    &= \sqrt{(l_L+2)(l_L+1)l_L(l_L-1)}\, \tensor[_{0}]{Y}{_{l_Lm_L}}, \notag
\end{align}
\end{subequations}
we obtain
\begin{align}
    &2\bar{\Gamma}^4\Sigma S_4^{(2)} \times\frac{16}{\Delta^6} = (l_L+2)(l_L-1)\tensor[_{-1}]{{Y^2_{l_Lm_L}}}{} \,A_{1}(r) \notag \\
    &+ \sqrt{(l_L+2)(l_L+1)l_L(l_L-1)}\,\, \tensor[_{-2}]{Y}{_{l_Lm_L}}\times\tensor[_{0}]{Y}{_{l_Lm_L}} A_{2}(r)
\end{align}
Since $\Gamma=\bar{\Gamma}=r$, $A_1$ and $A_2$ are now independent of $\theta$.

\begin{figure*}[htb]
   \centering
    \subfloat[$(2,2,0)\times(2,2,0)\to(5,4)$\label{fig:angular_projection_220_540}]{\includegraphics[width=0.5\textwidth,clip=true]{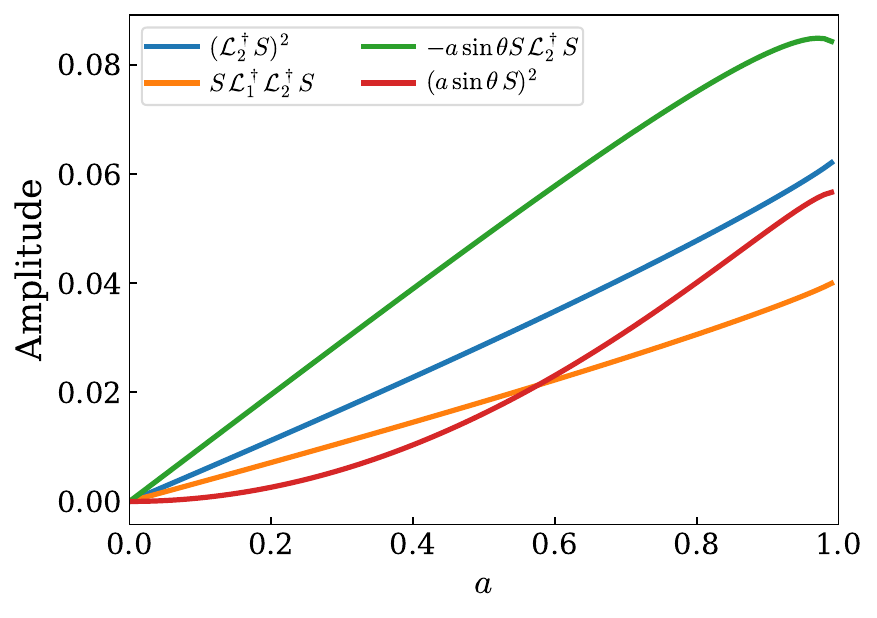}
    \includegraphics[width=0.5\textwidth,clip=true]{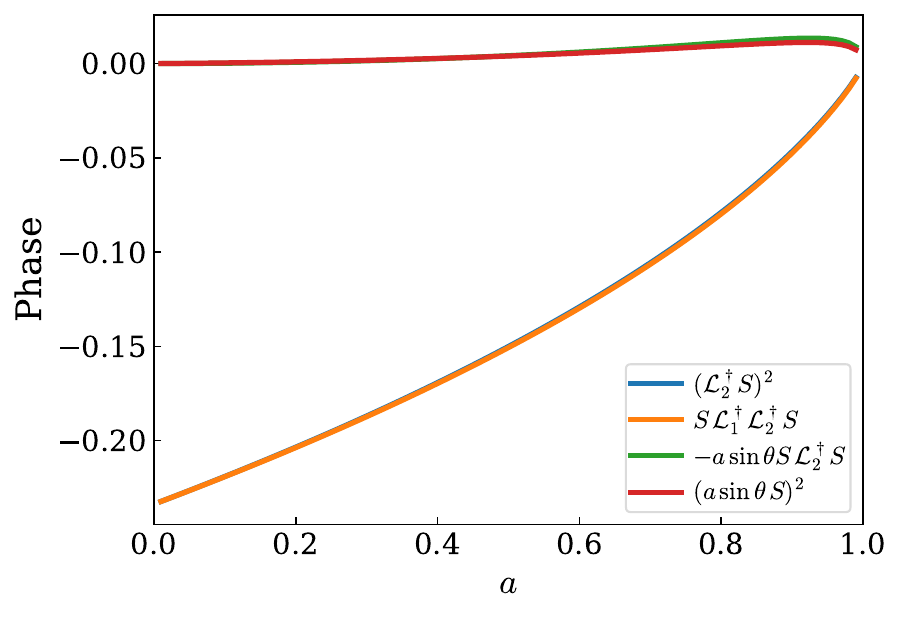}} \\
    \subfloat[$(2,2,0)\times(2,2,0)\to(4,4)$\label{fig:angular_projection_220_440}]{\includegraphics[width=0.5\textwidth,clip=true]{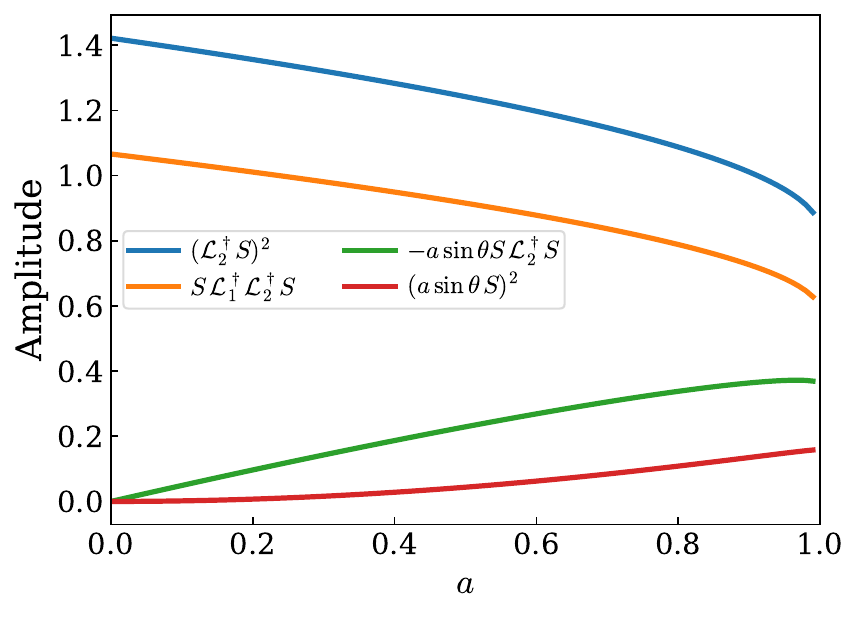}
    \includegraphics[width=0.5\textwidth,clip=true]{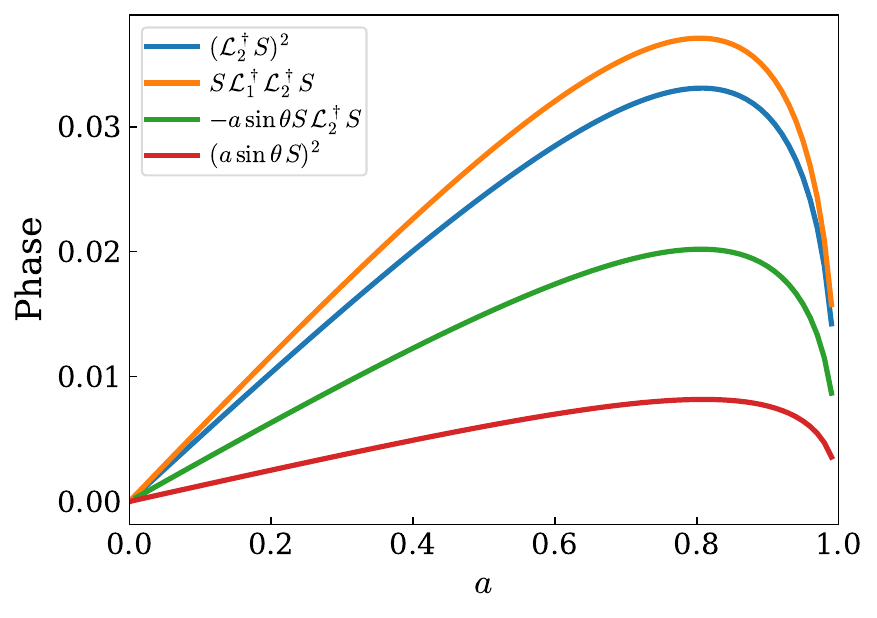}}
  \caption{The spin dependence of angular inner products between the spin-weighted spheroidal harmonic for the quadratic mode  $S_{l_{Q}m_Q}$ and four angular bases $\left(\mathcal{L}_2^\dagger S_{l_Lm_L}\right)^2$ (blue), $S_{l_Lm_L}\,\mathcal{L}_1^\dagger\mathcal{L}_2^\dagger S_{l_Lm_L}$ (orange), $-a\sin\theta\,S_{l_Lm_L} \mathcal{L}_2^\dagger S_{l_Lm_L}$ (green), and $\left(a\sin\theta S_{l_Lm_L}\right)^2$ (red). Here we consider $l_L=m_L=2$, and $(l_{Q}=5,m_{Q}=4)$ (top row) and $l_{Q}=m_{Q}=4$ (bottom row). The right (left) column corresponds to the amplitude (phase) of the inner products. The spin ranges from 0 to 0.99.}
\end{figure*}

We can see there are only two types of angular dependences: $\tensor[_{-1}]{{Y^2_{l_Lm_L}}}{}$, contributed by [Eq.~\eqref{eq:final_source_from_NR_variables}]
\begin{align*}
    (\bm{\bar{\delta}}+2\alpha+4\pi)^{(1)} \Psi_3^{(1)},
\end{align*}
and $\tensor[_{-2}]{Y}{_{l_Lm_L}}\times\tensor[_{0}]{Y}{_{l_Lm_L}}$, by
\begin{align*}
    -(\bm{D}+4\epsilon-\rho)^{(1)} \Psi_4^{(1)} -3\lambda^{(1)}\Psi_2^{(1)}.
\end{align*}
With this at hand, the angular projection in Eq.~\eqref{eq:2nd_equation_with_angular_projection} can be done analytically:
\begin{align}
    &\int \tensor[_{-2}]{Y}{_{l_Lm_L}}\times\tensor[_{0}]{Y}{_{l_Lm_L}}\times\tensor[_{-2}]{\bar{Y}}{_{l_Qm_Q}}   d\Omega \notag    \\
    &=\int (-1)^{m_Q}\tensor[_{-2}]{Y}{_{l_Lm_L}}\times\tensor[_{0}]{Y}{_{l_Lm_L}}\times\tensor[_{2}]{Y}{_{l_Q,-m_Q}}   d\Omega \notag \\
    &=(-1)^{m_Q}\sqrt{\frac{(2l_L+1)^2(2l_Q+1)}{4\pi}} \notag \\
    &\times
    \begin{pmatrix}
        l_L & l_L & l_Q \\
        m_L & m_L & -m_Q
    \end{pmatrix}
    \begin{pmatrix}
        l_L & l_L & l_Q \\
        2 & 0 & -2
    \end{pmatrix}, 
    \label{eq:3j_symbol_A}
\end{align}
and
\begin{align}
    &\int \tensor[_{-1}]{Y}{_{l_Lm_L}}^2\times\tensor[_{-2}]{\bar{Y}}{_{l_Qm_Q}}   d\Omega \notag \\
    &=    \int (-1)^{m_Q}\tensor[_{-1}]{Y}{_{l_Lm_L}}^2\times\tensor[_{2}]{Y}{_{l_Q,-m_Q}}   d\Omega \notag \\
    &=(-1)^{m_Q}\sqrt{\frac{(2l_L+1)^2(2l_Q+1)}{4\pi}} \notag \\
    &\times
    \begin{pmatrix}
        l_L & l_L & l_Q \\
        m_L & m_L & -m_Q
    \end{pmatrix}
    \begin{pmatrix}
        l_L & l_L & l_Q \\
        1 & 1 & -2
    \end{pmatrix},
    \label{eq:3j_symbol_B}
\end{align}
where we have used the $3-j$ symbol. The corresponding selection rule reads
\begin{align}
    &2m_L=m_Q,
    &l_Q\leq 2l_L. \label{eq:selection_rule_schwarzschild}
\end{align}
For $l_L=m_L=2$ and $l_Q=m_Q=4$, Eq.~\eqref{eq:3j_symbol_A}$=\frac{5}{2\sqrt{42\pi}}$, Eq.~\eqref{eq:3j_symbol_B}$=\frac{5}{3\sqrt{7\pi}}$.

As discussed in Appendix B of  \cite{Lagos:2022otp}, in addition to $\tensor[_{-1}]{{Y^2_{l_Lm_L}}}{}$ and $\tensor[_{-2}]{Y}{_{l_Lm_L}}\times\tensor[_{0}]{Y}{_{l_Lm_L}}$, the other two spin-weighted spherical harmonic multiplications: $\tensor[_{-3}]{Y}{_{l_Lm_L}}\times\tensor[_{1}]{Y}{_{l_Lm_L}}$ and $\tensor[_{-4}]{Y}{_{l_Lm_L}}\times\tensor[_{2}]{Y}{_{l_Lm_L}}$, also have nontrivial angular overlaps with the child mode $\tensor[_{-2}]{Y}{_{l_Qm_Q}}$, this can be seen from
\begin{align}
    &\int \tensor[_{s_1}]{Y}{_{l_Lm_L}}\times\tensor[_{s_2}]{Y}{_{l_Lm_L}}\times\tensor[_{-2}]{\bar{Y}}{_{l_Qm_Q}}   d\Omega \notag \\
    &\sim 
    \begin{pmatrix}
        l_L & l_L & l_Q \\
        s_1 & s_2 & -2
    \end{pmatrix},
\end{align}
which requires $s_1+s_2=-2$. However, since the spin weights of all the NP variables range from $-2$ to $+2$, there are no available scalars to be coupled with spin 1 (e.g. $\Psi_1^{(1)}$) and 2 (e.g. $\Psi_0^{(1)}$) fields at the second order. Therefore, these two angular components do not contribute to the equation. In fact, $\Psi_1$ and $\Psi_0$ never formally enter into any order of perturbation equations \cite{Campanelli:1998jv}. 

\subsection{Kerr black holes}
\label{subsec:angular_projection_Kerr_BH}
For Kerr BHs, the angular projection is complicated by the coefficients $A_{1..4}$ in Eq.~\eqref{eq:final_source_term}, which gain $\theta$-dependence via $\Gamma$ and $\bar{\Gamma}$. To construct an intuition into its spin dependence, here we simply consider the projection for the four angular bases
\begin{align}
    &\left<\left(\mathcal{L}_2^\dagger S_{l_Lm_L}\right)^2\right|\left.S_{l_{Q}m_Q}\right>,  \left<S_{l_Lm_L}\,\mathcal{L}_1^\dagger\mathcal{L}_2^\dagger S_{l_Lm_L}\right|\left.S_{l_{Q}m_Q}\right>, \notag \\
    &\left<a\sin\theta\,S_{l_Lm_L} \mathcal{L}_2^\dagger S_{l_Lm_L}\right|\left.S_{l_{Q}m_Q}\right>,  \notag \\
    &\left<\left(a\sin\theta S_{l_Lm_L}\right)^2\right|\left.S_{l_{Q}m_Q}\right>, \label{eq:four_angular_projections_kerr}
\end{align}
where the angular inner product $\left<\ldots\right|\left.\ldots\right>$ is defined in Eq.~\eqref{eq:angular_inner_product}.
As we will show below in Sec.~\ref{subsec:num_result}, the nature of the four integrals can offer a qualitative understanding of how the excitability of a quadratic QNM depends on spins.

Generally speaking, the coupling $(l_L,m_L,n_L)\times (l_L,m_L,n_L) \to (l_Q,m_Q)$ can be classified into two scenarios: when $l_Q> 2l_L$ and when $l_Q\leq 2l_L$. The former violates the selection rule in Eq.~\eqref{eq:selection_rule_schwarzschild} for Schwarzschild BHs, resulting in the absence of any quadratic excitation. However, with an increase in the spins of BHs, the selection rule is broken, leading to an amplification of the effect. For instance, Figure \ref{fig:angular_projection_220_540} shows the spin dependency of the integrals for $(l_L=m_L=2,n_L=0)^2 \to (l_Q=5,m_Q=4)$. In this calculation, we adopt the Black Hole Perturbation Toolkit \cite{BHPToolkit} for the evaluation of spin-weighted spheroidal harmonics and employ Gauss–Legendre quadrature for integration. The results reveal that the amplitudes of the angular integrals vanish when $a=0$ and demonstrate a nearly monotonic increase with spin. The term of $(a\sin\theta\,S_{l_Lm_L} \mathcal{L}_2^\dagger S_{l_Lm_L})$ offers the strongest angular contribution.

By contrast, the second scenario $l_Q\leq 2l_L$ satisfies the selection rule. When $a=0$, the first two integrals in Eq.~\eqref{eq:four_angular_projections_kerr} reduce to the $3-j$ symbols\footnote{They differ by a factor of $2\pi$ since the integration in Eq.~\eqref{eq:angular_inner_product} does not include the azimuthal part.}. Figure \ref{fig:angular_projection_220_440} displays the results for $(l_L=m_L=2,n_L=0)^2 \to (l_Q=m_Q=4)$. We can see the dominant components $\left(\mathcal{L}_2^\dagger S_{l_Lm_L}\right)^2$ and $(S_{l_Lm_L}\,\mathcal{L}_1^\dagger\mathcal{L}_2^\dagger S_{l_Lm_L})$ decrease with spin.

\section{Dealing with divergence at boundaries: the complex plane approach}
\label{sec:contour}
For both linear and nonlinear QNM analysis, necessary regularization procedures are often required to deal with the ``blowing-up'' wavefunctions at spatial infinity and BH horizon\footnote{Another approach is to solve the wave on a hyperboloidal slicing, as implemented in \cite{Ripley:2020xby}.}. For example, Detweiler \cite{Detweiler:1979xr} introduced counter-terms to eliminate singular terms for QNM excitation computations, while Leaver \cite{Leaver:1986gd} proposed integrating along contours in the complex$-r$ domain for similar problems. Later, the complex-contour technique was used to introduce a mode ``inner-product'' or ``bi-linear form'', which formed the basis for the investigation of linear and nonlinear waves in Kerr/Kerr-Newman spacetimes \cite{Yang:2014tla, Mark:2014aja, Zimmerman:2014aha}, as well as exploring the spectral stability of near-extremal spacetimes \cite{Yang:2022wlm}. Recently, a new bi-linear form was motivated by the ``conserved current'' associated with the Teukolsky operator \cite{Green:2022htq}, and it was employed to derive QNM orthogonality.

In this work, we adopt this complex-plane technique to solve the second-order equation in Eq.~\eqref{eq:2nd_equation_with_angular_projection}. Below, we first provide some basic information about the method in Sec.~\ref{subsec:contour_review}. Next in Sec.~\ref{subsec:bi_linear_form}, we review the idea of the ``bi-linear form'' defined based on the complex-plane technique. In particular, we explain how it facilitates the studies of eigenvalue perturbation (Sec.~\ref{subsubsec:eigenvalue_perturbation}) and QNM orthogonality (Sec.~\ref{subsubsec:qnm_orthogonality}). We will show that the two bi-linear forms used in Refs.~\cite{Yang:2014tla, Mark:2014aja, Zimmerman:2014aha} and \cite{Green:2022htq} yield the same QNM orthogonality conclusion. At last, in Sec.~\ref{subsec:contour_2nd_excitation}, we explain how to implement the complex-contour for computing second-order QNMs as the main purpose of this work.

\begin{figure}[tb]
\begin{tikzpicture}
  \draw[thick, ->, >=stealth] (-1,0) -- (6.5,0) node[right] {Re $r$};
  \draw[thick, ->, >=stealth] (0,-1.5) -- (0,5) node[above] {Im $r$};

  \draw (4.5,-0.1) -- (4.5,0.1) node[below] at (4.6,-0.1) {\normalsize $r_+$};
  \draw (2.1,-0.1) -- (2.1,0.1) node[below] at (2.2,-0.1) {\normalsize $r_-$};

   \draw[decorate, decoration={zigzag, segment length=4, amplitude=1, pre=lineto, post=lineto}, red, line width=1.5pt] (4.5,0) -- (4.5,5);

   \fill[red] (4.5,0) circle[radius=2pt];

   \draw[blue, line width=1.5pt, decoration={markings, mark=at position 0.5 with {\arrow[scale=2]{>}}}, postaction={decorate}] (3.3,5) -- (3.3,-1.1) node[midway,left,left=3mm]{L};
   \draw[blue, line width=1.5pt, decoration={markings, mark=at position 0.6 with {\arrow[scale=2]{>}}}, postaction={decorate}] (5.9,-1.1) -- (3.3,-1.1) node[midway,below,below=3mm]{B};
   \draw[blue, line width=1.5pt, decoration={markings, mark=at position 0.5 with {\arrow[scale=2]{>}}}, postaction={decorate}] (5.9,5) -- (5.9,-1.1) node[midway,right,right=3mm]{R};

   \fill[YellowOrange] (3.32,-1.1) circle[radius=2pt] ;

   \node[below] at (3.3,-1.1) {\parbox{1.5cm}{matching \\ point}};

   \node[below] at (2.9,4) {\large $\mathcal{C}$};

   \fill[YellowOrange] (5.9,5) circle[radius=2pt];
   \node[right] at (5.91,4.95) {a};

   \fill[YellowOrange] (3.3,5) circle[radius=2pt];
   \node[right] at (3.3,4.95) {b};

\end{tikzpicture}
\caption{The contour (labeled by $\mathcal{C}$, in blue) and branch cut (in red) used in this work.  The contour goes around the $r_+$ singularity and the corresponding branch cut. There is a second branch cut connecting $r_-$ which is not shown in this figure. Three blue arrows indicate the stable integration direction while solving for QNM wavefunctions: Two solutions are shot separately from ``a'' and ``b'', and they are matched at the bottom left corner labeled by ``matching point''. In our numerical implementation, we fix the real part of the left and right vertical paths to $(r_-+r_+)/2=1$ and $3$, respectively. The imaginary part of the bottom path is fixed to $-5$.}
\label{fig:cp}
\end{figure}
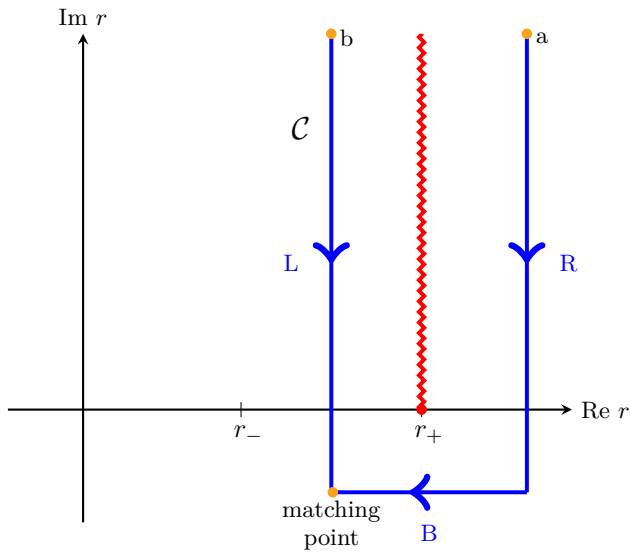

\subsection{Branch cut and contour}
\label{subsec:contour_review}
The frequency-domain Teukolsky equation is usually solved on the real$-r$ axis. However, it is also instructive to extend both the equation and its solutions in the complex$-r$ domain, in which case the analytical properties of the solution, e.g., singularities and branch cuts become important. For example, the scalar function $X_L(r)$ in Leaver's solution [Eqs.~\eqref{eq:leaver_method_s} and \eqref{eq:Leaver_XL}] has two singularities at $r_\pm$ and the branch cuts ending at them. In Fig.~\ref{fig:cp}, the zigzag red line illustrates a convenient choice for the branch cut. The corresponding logarithmic expressions are given by
\begin{align}
    \lim_{r\to\rm right} \ln r =\ln |r| + \frac{i\pi}{2}, \quad
    \lim_{r\to\rm left} \ln r =\ln |r| - \frac{3i\pi}{2}.
\end{align}
They capture the behavior of the logarithm as $r$ approaches the right and left sides of the cut, respectively.

Alongside the branch cut, we introduce a contour $\mathcal{C}$ in Fig.~\ref{fig:cp}. On both ends of this contour, the QNM boundary condition $\sim e^{i \omega_{lmn}r_*}$ exhibits \textit{exponential decay} due to the positive real part of $\omega_{lmn}$. In addition, as the path remains a finite distance from the horizon, it avoids the blowing-up feature at $r_+$. Consequently, the wavefunction of a QNM exhibits regular behavior throughout $\mathcal{C}$ and can be safely treated with numerical methods.

\subsection{Bi-linear form}
\label{subsec:bi_linear_form}
Consider two generic wave functions $\chi(r,\theta),\eta(r,\theta)$ and  the frequency-domain Teukolsky opeartor $\mathcal{T}(\omega, m,r,\theta)$ [see Eq.~\eqref{eq:Teukolsky_operator_t_phi}, with $\partial/\partial t$ and $\partial/\partial \phi$ replaced by $-i\omega$ and $im$, respectively.]. A bi-linear form $\langle | \rangle$ can be defined on the contour $\mathcal{C}$
such that $\langle \chi | \mathcal{T} \eta\rangle =\langle  \mathcal{T \chi | \eta\rangle }$ \cite{Yang:2014tla, Mark:2014aja, Zimmerman:2014aha}:
\begin{align}\label{eq:bil}
\langle \chi | \eta \rangle & =\int_{\mathcal{C}} (r-r_+)^s(r-r_-)^s dr \int \sin\theta d \theta \,\chi(r,\theta) \eta(r,\theta)\nonumber \\
& = \langle \eta | \chi \rangle\,,
\end{align}
where $s$ is the spin of the field. If $\chi,\eta$ satisfy the QNM boundary conditions then the contribution from both ends of $\mathcal{C}$ may be exponentially suppressed similar to the QNMs.

\subsubsection{Eigenvalue perturbation}
\label{subsubsec:eigenvalue_perturbation}
The bi-linear form can be used to study the eigenvalue perturbation problem. Let us consider a QNM wavefunction satisfying $\mathcal{T}(\omega_0) \psi_0=0$ with a certain eigenfrequency $\omega_0$. If the wave equation is modified, e.g. due to the presence of additional fields \cite{Mark:2014aja} and/or modification of the underlying gravity theory \cite{Li:2022pcy,Hussain:2022ins,Wagle:2023fwl}, the wave operator can often be written as $\mathcal{T} +\epsilon\Delta \mathcal{T}$ with $\epsilon$ being a book-keeping index, and the corresponding new eigenmode satisfies
\begin{align}
\left [\mathcal{T}(\omega_0+\epsilon \omega_1)+\epsilon\Delta \mathcal{T}(\omega_0+\epsilon \omega_1) \right ] (\psi_0+\epsilon\psi_1)=0\,,
\end{align}
where both the eigenfrequency and mode wavefunction are expanded up to the $\mathcal{O}(\epsilon)$ order. Up to the same order in $\epsilon$ the above equation may be rewritten as
\begin{align}\label{eq:1sto}
\mathcal{T}(\omega_0) \psi_1+\omega_1 \partial_\omega \mathcal{T} \psi_0+\Delta \mathcal{T}(\omega_0) \psi_0=0\,.
\end{align}

With the bi-linear form introduced in Eq.~\eqref{eq:bil} we can realize that
\begin{align}
\langle \psi_0 | \mathcal{T}(\omega_0) \psi_1 \rangle =\langle \mathcal{T}(\omega_0) \psi_0 | \psi_1 \rangle =0
\end{align}
so that Eq.~\eqref{eq:1sto} can be used to solve for $\omega_1$ without knowing the expression of $\psi_1$:
\begin{align}
\omega_1 =-\frac{\langle \psi_0 | \Delta \mathcal{T} \psi_0 \rangle}{\langle \psi_0 | \partial_\omega \mathcal{T} \psi_0 \rangle}\,,
\end{align}
which is similar to the perturbation theory for non-degenerate stationary states in quantum mechanics. See also \cite{Cannizzaro:2023jle} for a relevant discussion on BH boson clouds.

\subsubsection{Quasinormal mode orthogonality}
\label{subsubsec:qnm_orthogonality}
The ``self-jointness'' of the bi-linear form $\langle \chi | \mathcal{T} \eta\rangle =\langle  \mathcal{T \chi | \eta\rangle }$ provides an alternative rationale for the orthogonality of QNMs, supplementing the methods provided in \cite{Green:2022htq}.
Let us now consider two QNMs satisfying $\mathcal{T}(\omega_a) \psi_a=0, \mathcal{T}(\omega_b) \psi_b=0$ respectively, assuming  azimuthal number $m_a$ and $m_b$. As a result, we expect that
\begin{align}
\langle \psi_b | \mathcal{T}(\omega_a) \psi_a \rangle =0, \quad \langle \psi_a | \mathcal{T}(\omega_b) \psi_b \rangle =0
\end{align}
According to the properties of the bi-linear form, the above equations imply that
\begin{align}
\langle \psi_a | (\mathcal{T}(\omega_a)-\mathcal{T}(\omega_b)) \psi_b \rangle =0\,.
\end{align}
This equation can be interpreted as an orthogonal relation between the two mode wavefunctions $\psi_a, \psi_b$ with the weight function $\mathcal{T}(\omega_a)-\mathcal{T}(\omega_b)$ given by (choosing $m_a=m_b=m$)
\begin{widetext} 
\begin{align}
\mathcal{T}(\omega_a)-\mathcal{T}(\omega_b) =(\omega_a-\omega_b) \left \{-(\omega_a+\omega_b) \left [ \frac{(r^2+a^2)^2}{\Delta}-a^2\sin^2\theta\right ] +\frac{4  m a r}{\Delta }+2 i s  \left [-r-i a \cos\theta + \frac{r^2-a^2}{\Delta}\right ]\right \},
\end{align}
\end{widetext}
which is the same as Eq.~$(46)$ in \cite{Green:2022htq} by setting $s=-2$ and re-normalizing the constant factor that contains $\omega_a-\omega_b$\footnote{There are typos in their Eq.~$(55)$.}. 

\subsection{Second-order mode excitation}
\label{subsec:contour_2nd_excitation}
We now adopt the complex-contour technique to solve the second-order Teukolsky equation in Eq.~\eqref{eq:2nd_equation_with_angular_projection}. Using the contour $\mathcal{C}$ in Fig.~\ref{fig:cp}, it is convenient to decompose $r$ into its real and imaginary parts, $r=x+iy$. This decomposition allows us to transform Eq.~\eqref{eq:2nd_equation_with_angular_projection} into forms that are more amenable for numerical implementation on each segment of the contour. More specifically, along the vertical paths (labeled by ``L'' and ``R''), the equation reads 
\begin{align}
    & \frac{d^2 R_{l_Q}^{(2)}}{dy^2}=-\frac{Q(r)}{\Delta}+2i\frac{(r-1)}{\Delta}\frac{d R_{l_Q}^{(2)}}{dy}\notag \\
    &+\frac{R_{l_Q}^{(2)}}{\Delta}\left(\frac{K_Q^2+4iK_Q(r-1)}{\Delta}-16i\omega_L r-\lambda^{(2)}_{l_Q}\right), \label{eq:2nd_t_eq_vertical}
\end{align}
whereas on the horizontal path (labeled by ``B''), we have
\begin{align}
    & \frac{d^2 R_{l_Q}^{(2)}}{dx^2}=\frac{Q(r)}{\Delta}+2\frac{(r-1)}{\Delta}\frac{d R_{l_Q}^{(2)}}{dx}\notag \\
    &-\frac{R_{l_Q}^{(2)}}{\Delta}\left(\frac{K_Q^2+4iK_Q(r-1)}{\Delta}-16i\omega_L r-\lambda^{(2)}_{l_Q}\right). \label{eq:2nd_t_eq_horizontal}
\end{align}
The solution and its derivatives need to be continuous at corners, which yields continuity conditions:
\begin{subequations}
\label{eq:continuity_eq}
    \begin{align}
    &\left.R_{l_Q}^{(2)}\right|_{\rm{horizontal}}=\left.R_{l_Q}^{(2)}\right|_{\rm{vertical}}, \\
    &\left.\frac{d}{dx}R_{l_Q}^{(2)}\right|_{\rm{horizontal}}=\left.-i\frac{d}{dy}R_{l_Q}^{(2)}\right|_{\rm{vertical}}. 
\end{align}
\end{subequations}

To impose boundary conditions at ``a'' and ``b'' in Fig.~\ref{fig:cp}, we can use the asymptotic expansion of $R_{l_Q}^{(2)}$ in Eq.~\eqref{eq:RQ_asymptotic}. It leads to
\begin{subequations}
\label{eq:bc_condition}
\begin{align}
    R_{l_Q}^{(2)}&=E \left(1+\frac{a_Q}{r}\right) (r-r_+)\Delta X_Q  \label{eq:2ndQNM_bc_R}, \\
    \frac{d}{dy}R_{l_Q}^{(2)}&=iE \left(1+\frac{a_Q}{r}\right) \frac{d}{d r} (r-r_+)\Delta X_Q,
\end{align}
\end{subequations}
where $X_Q=X_L^2$ and $X_L$ is defined in Eq.~\eqref{eq:Leaver_XL}. The factor $(1+a_Q/r)$ shows that the conditions above are accurate up to $\mathcal{O}(r^{-1})$.

The unknown constant number $E$ corresponds to the amplitude of the quadratic QNM with frequency $2\omega_L$. To see this, we adopt the relation 
\begin{align}
   \bar{\Gamma}^4\Psi_4^{(2)}= e^{-2i\omega_{L} t} R^{(2)}_{l_Q}(r) \, S^{(2)}_{l_Q}(\theta)e^{2im_L \phi}. \notag
\end{align}
After plugging Eq.~\eqref{eq:2ndQNM_bc_R} into the above expression, the asymptotic behavior of the second perturbation $\Psi_4^{(2)}$ is given by
\begin{align}
    \left[r\Psi_4^{(2)}\right]_{l_Q,m_Q=2m_L}^{(\infty)}=E e^{-2i\omega_{L} (t-r_*)}.
\end{align}
Meanwhile, the amplitude of the linear perturbation $\Psi_4^{(1)}$ is provided in Eq.~\eqref{eq:linear_psi4_inf}. Their ratio, namely the excitation factor of the quadratic QNM with frequency $2\omega_L$, reads
\begin{align}
    &M_{\Psi_4}^{\rm Kinnersley} = \frac{\left[r\Psi_4^{(2)}\right]_{l_Q,m_Q}^{(\infty)}}{\left\{\left[r\Psi_4^{(1)}\right]_{l_L,m_L}^{(\infty)}\right\}^2}=\frac{4E}{\omega_{L}^8(\sum a_n)^2}. \label{eq:excitation_factor_psi4_kinn}
\end{align}
We note that the value of the excitation factor for $\Psi_4$  relies on the choice of tetrad. The result in Eq.~\eqref{eq:excitation_factor_psi4_kinn} is associated with the Kinnersley tetrad in Eq.~\eqref{eq:Kinnersley}. It differs from the one used by the numerical-relativity code SpEC \cite{SpECwebsite,Iozzo:2020jcu}:
\begin{align}
    \bm{l}^{\rm SpEC}=\frac{1}{\sqrt{2}}(\partial_t+\partial_r), \quad
    \bm{n}^{\rm SpEC}=\frac{1}{\sqrt{2}}(\partial_t-\partial_r).
\end{align}
The conversion is given by \cite{Zhu:2024rej}
\begin{align}
    M_{\Psi_4}^{\rm SpEC}=\frac{1}{2}M_{\Psi_4}^{\rm Kinnersley}.
\end{align}
Furthermore, the excitation factor for the strain can be obtained straightforwardly via the relation $\Psi_4^{\rm SpEC}=-\ddot{h}$, which yields
\begin{align}
    &M_{h}^{\rm SpEC}=\frac{\omega_L^2}{4}M_{\Psi_4}^{\rm SpEC} =\frac{E}{2\omega_{L}^6(\sum a_n)^2}. \label{eq:excitation_factor_h_spec}
\end{align}

Therefore, to compute the excitation factor $M_{h}^{\rm SpEC}$, it suffices to integrate Eqs.~\eqref{eq:2nd_t_eq_vertical} and \eqref{eq:2nd_t_eq_horizontal} along with the continuity conditions in Eqs.~\eqref{eq:continuity_eq} and the boundary conditions in Eqs.~\eqref{eq:bc_condition}. It is important to note that there exists a stable integration direction on the vertical paths. This is because Eq.~\eqref{eq:2nd_t_eq_vertical} admits two independent solutions. One corresponds to the QNM solution, asymptoting to $e^{2i\omega_L r_*}\sim e^{-2\Re \omega_L y}$ as $y\to \infty$. The second one is non-physical, diverging at infinity $ e^{-2i\omega_L r_*}\sim e^{2\Re \omega_L y}$. At the initial point of the integration, due to finite numerical accuracy, the non-physical solution emerges alongside the physical one. To ensure stability, an integration direction must be chosen such that the non-physical solution decays throughout the integration, preventing it from exponentially blowing up and eventually overshadowing the physical QNM solution. Consequently, the stable direction on both vertical paths has to be top-down. In practice,  the shooting method is employed to achieve this goal: using the explicit Runge-Kutta method of order 5, two solutions are shot separately from ``a'' and ``b'', with the same shooting parameter $E$. At the matching point in Fig.~\ref{fig:cp}, the Nelder–Mead method is adopted to match the two shoots through the continuity conditions in Eq.~\eqref{eq:continuity_eq}.

\section{Numerical calculations and results}
\label{sec:results}
Having discussed two basic ingredients: metric reconstruction and the complex-plane technique, we are now in the position to compute the value of the excitation factor for quadratic QNMs. To start with, we explore a toy model in Sec.~\ref{subsec:num_toy_model} to establish a qualitative understanding of the numerical aspects for the complex-plane method. In Sec.~\ref{subsec:num_result}, we proceed to investigate quadratic QNMs.

\subsection{A toy model}
\label{subsec:num_toy_model}
We consider a source $Q(r)$ that has a form of
\begin{align}
    Q(r)=X_Q (r-r_+)  \Delta\sum_{n=0} f_n\left(\frac{r-r_+}{r-r_-}\right)^n.
\end{align}
Following \cite{Nakano:2007cj}, we use Leaver's representation to construct the solution \cite{leaver1985analytic}
\begin{align}
    R_{l_Q}^{(2)}(r) =  X_Q (r-r_+) \Delta \sum_{n=0} d_n\left(\frac{r-r_+}{r-r_-}\right)^n .
\end{align}
By virtue of the Teukolsky equation, the coefficients $d_n$ are determined by three-term recurrence relations
\begin{subequations}
\begin{align}
    & \alpha_0 d_1 +\beta_0 d_0 = f_0, \\
    &\alpha_n d_{n+1}+\beta_n d_n+\gamma_n d_{n-1}=f_n.
\end{align} 
\end{subequations}
where $\alpha_n$, $\beta_n$, and $\gamma_n$ are given in Eq.~\eqref{eq:leaver_alpha_beta_gamma}. We consider a special source: 
\begin{align}
    f_0=\alpha_0+\beta_0,\,\, f_1=\beta_1+\gamma_1, \,\, f_2=\gamma_2, \, \, f_{n>2}=0. \notag
\end{align}
The corresponding solution has an analytic expression
\begin{align}
    R_{l_Q}^{(2)}(r) =  X_Q \Delta (r-r_+)\left(1+\frac{r-r_+}{r-r_-}\right). \label{eq:toy_model_analytic_solution}
\end{align}
This implies that $E=2$ and $a_Q=-\sqrt{1-a^2}$ in Eq.~\eqref{eq:bc_condition}.

\begin{figure}[htb]
   \centering
    \includegraphics[width=0.5\textwidth,clip=true]{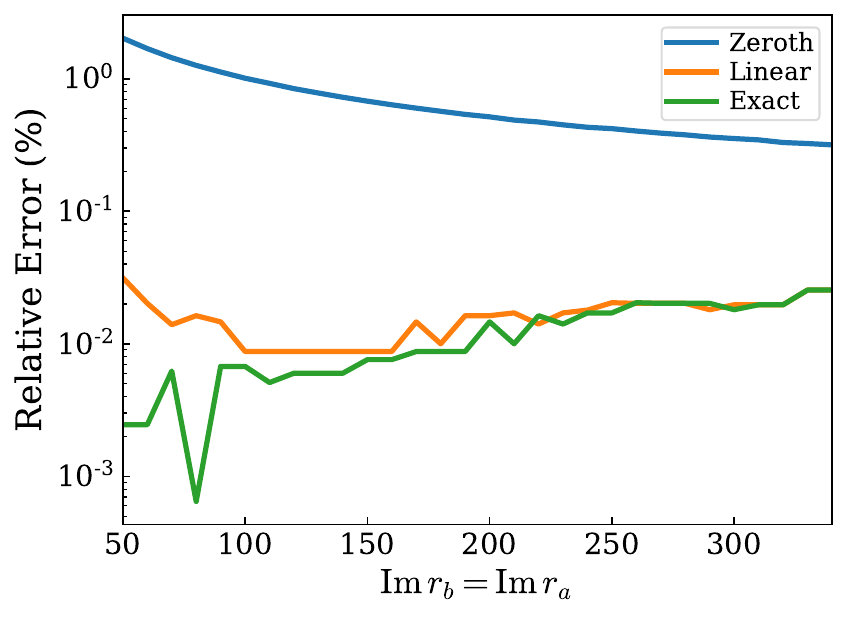}
  \caption{The relative error of the computed $E$ in Eq.~\eqref{eq:bc_condition} compared to the expected value of 2 for the toy model considered in Sec.~\ref{subsec:num_toy_model}. The boundary condition is accurate to the zeroth (in blue) and first order (in orange). The green curve corresponds to the exact boundary condition.  }
 \label{fig:toy_model_radius}
\end{figure}

Then we use the method given in Sec.~\ref{subsec:contour_2nd_excitation} to compute the value of $E$ numerically. Our numerical settings are listed below
\begin{align}
    {\rm Re}\, r_{\rm b} =\frac{r_++r_-}{2}=1, \quad {\rm Re}\, r_{\rm a}=3, \quad {\rm Im}\, r_{\rm matching} = -5, \label{eq:contour_corner_infos}
\end{align}
where the subscripts ``a'', ``b'', and ``matching'' refer to three points in Fig.~\ref{fig:cp}. We have checked that varying their values introduces a relative error of less than $10^{-3}\%$ in the results. The choice of the heights of the vertical paths, namely ${\rm Im}\, r_{\rm a}$ and ${\rm Im}\, r_{\rm b}$, requires more attention. For simplicity, we set both heights to the same number.
On the one hand, positioning the boundary at small distances can result in non-negligible systematic errors as the boundary conditions in Eqs.~\eqref{eq:bc_condition} are accurate only to linear order. On the other hand, since the solution decays exponentially at a large distance, placing the boundary too far may cause the integrator to lose precision, especially when the wavefunction spans many orders of magnitude. To see this, we calculate the value of $E$ for the $(l=m=2,n=0)$ mode of a Schwarzschild BH at a variety of heights. The orange curve in Fig.~\ref{fig:toy_model_radius} displays its deviation from the theoretical value 2. We can see the optimum window is around $[100, 160]$, beyond which the error gradually worsens as the boundary distance further increases. Nevertheless, the relative error remains below $3\times10^{-2}\,\%$ throughout the entire range.

\begin{figure}[htb]
   \centering
    \includegraphics[width=0.5\textwidth,clip=true]{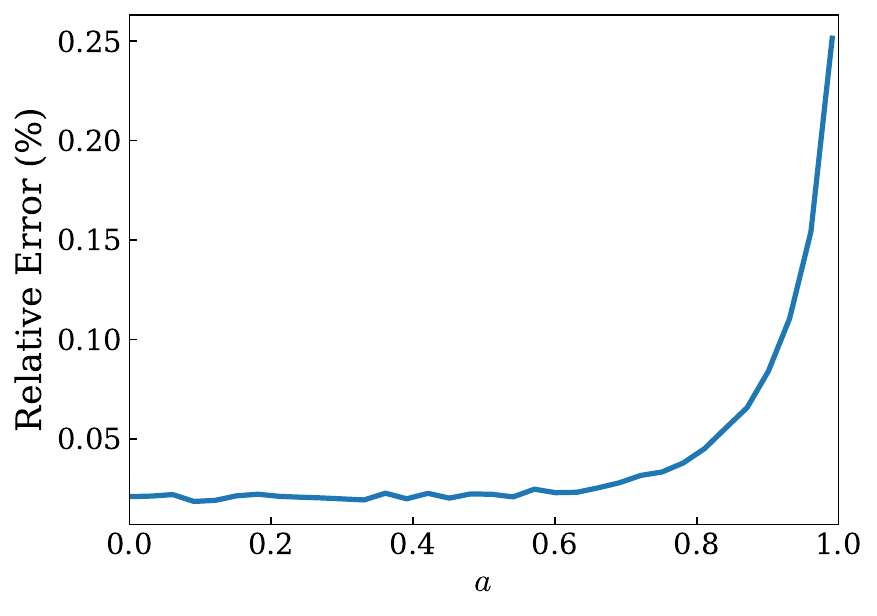}
  \caption{The spin dependence of the relative error for the computed $E$ compared to the expected value of 2, using the toy model considered in Sec.~\ref{subsec:num_toy_model}. We set ${\rm Im}\, r_{\rm a}={\rm Im}\, r_{\rm b}=100$, and the boundary condition is accurate to the first order.}
 \label{fig:toy_model_spin}
\end{figure}

Since the analytic expression of the solution is available in Eq.~\eqref{eq:toy_model_analytic_solution}, we can refine the boundary conditions to incorporate their exact expressions. The result is shown as the green curve in Fig.~\ref{fig:toy_model_radius}. When the height of the vertical paths is larger than 100, the error of $E$ gradually approaches the linear-order result at greater distances, suggesting that the precision of the integrator becomes the primary source of inaccuracy.

To make a more comprehensive comparison, we also demote the boundary conditions in Eqs.~\eqref{eq:bc_condition}  to zeroth order by setting $a_Q$ to 0. The blue curve in Fig.~\ref{fig:toy_model_radius} shows the corresponding result. This time, the relative errors increase to $\sim 1\%$. We find the curve can be well-fitted with a $1/r$ form, suggesting the necessity of the linear-order correction in our studies.

Finally, we switch our attention to the effect of BH's spin. We fix ${\rm Im}\, r_{\rm a}$ and ${\rm Im}\, r_{\rm b}$ at 100, and vary the spin $a$ from 0 to 0.99. As shown in Fig.~\ref{fig:toy_model_spin}, the error increases with $a$, particularly when $a>0.8$. This trend can be attributed to the rising real part of the complex frequency with increasing values of $a$, leading to a faster decay of the wavefunction along the vertical paths and resulting in larger numerical errors within the integrator. Nevertheless, the relative error is still less than $0.25\%$ at $a=0.99$.

\begin{figure*}[htb]
   \centering
    \subfloat[$(2,2,0)\times(2,2,0)\to(4,4)$\label{fig:excitation_factor_220_440}]{\includegraphics[width=0.5\textwidth,clip=true]{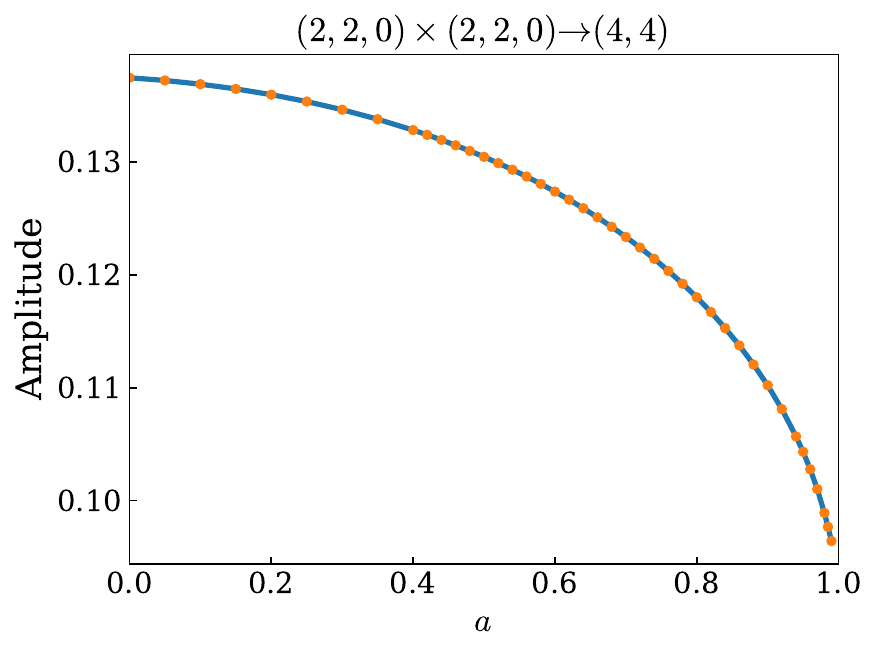}
    \includegraphics[width=0.5\textwidth,clip=true]{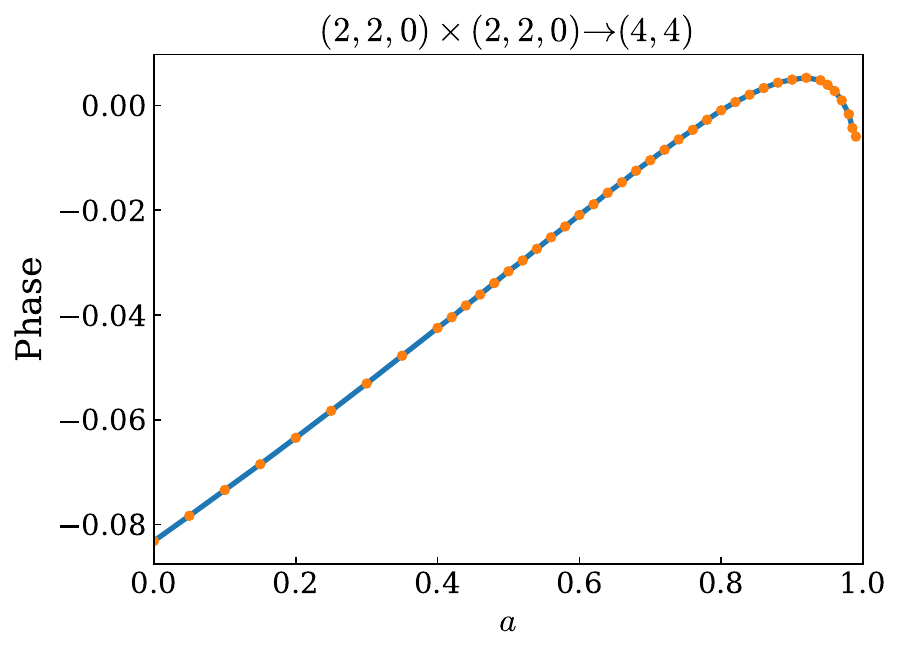}} \\
    \subfloat[$(2,2,0)\times(2,2,0)\to(5,4)$\label{fig:excitation_factor_220_540}]{\includegraphics[width=0.5\textwidth,clip=true]{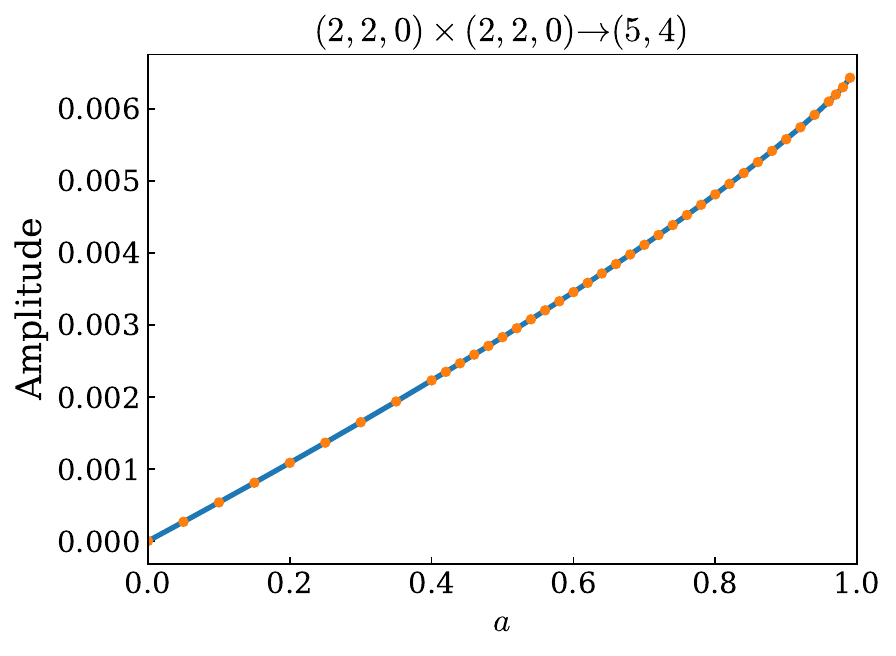}
    \includegraphics[width=0.5\textwidth,clip=true]{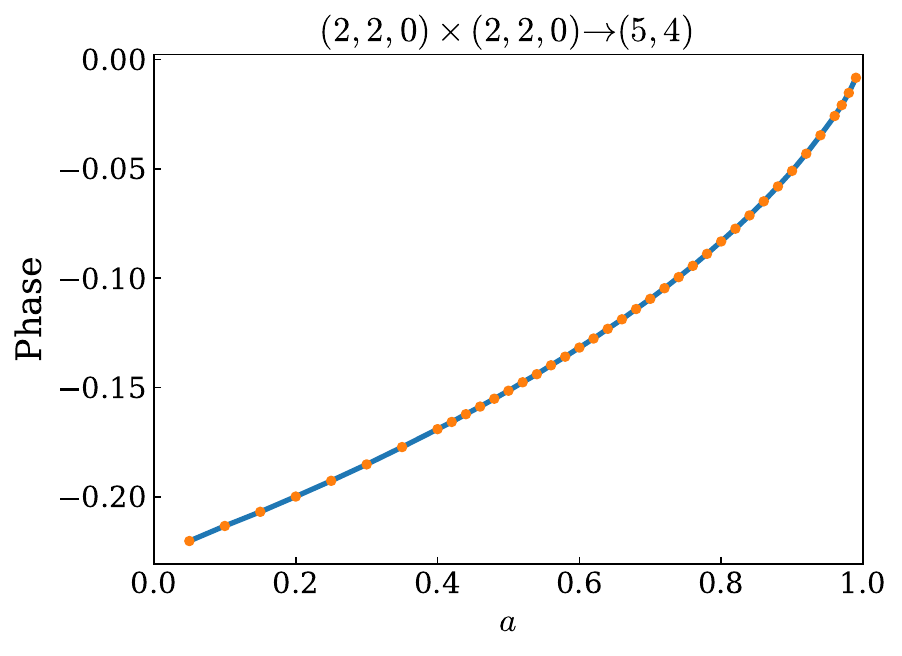}}
  \caption{The spin dependence of the excitation factor $M_{h}^{\rm SpEC}$ defined in Eq.~\eqref{eq:excitation_factor_h_spec}, for the channel $(2,2,0)\times(2,2,0)\to(4,4)$ (top) and $(2,2,0)\times(2,2,0)\to(5,4)$ (bottom). The right (left) column corresponds to its amplitude (phase). The spin ranges from 0 to 0.99. }
\end{figure*}

\subsection{Quadratic quasinormal modes}
\label{subsec:num_result}
 We are now ready to solve the quadratic Teukolsky equation in Eq.~\eqref{eq:2nd_equation_with_angular_projection}. The treatment for the left-hand-side (namely the integrator) is the same as the toy model in Sec.~\ref{subsec:num_toy_model}, as outlined in Sec.~\ref{subsec:contour_2nd_excitation}. The only difference is the source $Q(r)$ in Eq.~\eqref{eq:2nd_equation_with_angular_projection}. 

We first notice that $Q(r)$ can be decomposed into the following form 
\begin{align}
    &Q(r)= r^{12} \left(\tensor[_{+2}]{{R_{lm\omega_L}}}{}\right)^2 \times W_1(r) \notag \\
    &+ r^{10}\Delta\,\tensor[_{+2}]{{R_{lm\omega_L}}}{}\left(\frac{d}{dr}\tensor[_{+2}]{{R_{lm\omega}}}{}\right) \times W_2(r) \notag \\
    &+r^8\Delta^2\left(\frac{d}{dr}\tensor[_{+2}]{{R_{lm\omega_L}}}{}\right)^2 \times W_3(r), \label{eq:leaver_source_q_R_dr_R}
\end{align}
where the first-order Teukolsky equation has been used to repeatedly substitute $d^2\tensor[_{+2}]{{R_{lm\omega_L}}}{}/dr^2$ and higher-order radial derivatives with $d\tensor[_{+2}]{{R_{lm\omega_L}}}{}/dr$ and $\tensor[_{+2}]{{R_{lm\omega_L}}}{}$. We then evaluate their values via Leaver's solution in Eq.~\eqref{eq:leaver_hertz}.
Meanwhile, the three functions $W_{1,2,3}(r)$ are complicated rational functions of $r$. Note that we have separated the leading terms $r^{12},r^{10}$ and, $r^8$ that are singlar at infinity. Consequently, $W_{1,2,3}(r)$ behave regularly across the entire contour $\mathcal{C}$. In practice, we find that evaluating the analytical expressions of $W_{1,2,3}(r)$ is time-consuming, adversely affecting the numerical efficiency of the integrator. To address this issue, we adopt linear spline interpolation to obtain the values of the regular functions along $\mathcal{C}$. 

In this paper, we restrict ourselves to the quadratic effect of the $(l=m=2,n=0)$ QNM. Below we still use $\omega_L$ to represent its complex frequency. We compute its value using a Python package $\textsf{qnm}$ \cite{Stein:2019mop}. The corresponding separation constant $\lambda^{(2)}_{l_Q}$ in Eq.~\eqref{eq:2nd_equation_with_angular_projection}, which is associated with the quadratic mode $2\omega_L$, can be evaluated with the Black Hole Perturbation Toolkit \cite{BHPToolkit}. Following Sec.~\ref{subsec:num_toy_model}, our numerical settings for the contour $\mathcal{C}$ are the same as Eq.~\eqref{eq:contour_corner_infos}. Meanwhile, we set the height of vertical paths ${\rm Im}\, r_{\rm a}={\rm Im}\, r_{\rm b}$ to 100, which was shown to be suitable for the linear boundary conditions. We have checked that varying the height from 90 to 150 yields $\sim 0.1\%$ changes in results for Kerr BHs with various spins.

Figure \ref{fig:excitation_factor_220_440} displays the spin dependence of $M_{h}^{\rm SpEC}$, defined in Eq.~\eqref{eq:excitation_factor_h_spec}, for the excitation channel $(l_L=m_L=2,n_L=0)^2 \to (l_Q=m_Q=4)$. In the Schwarzschild case, the amplitude and phase of $M_{h}^{\rm SpEC}$ are 0.137 and $-0.083$, respectively. With the increase of the spin, the amplitude decreases monotonically. It approaches a value of 0.0964 when $a=0.99$. Meanwhile, the phase reaches its maximum value at $a\sim0.92$. Compared with Fig.~\ref{fig:angular_projection_220_440}, we find the spin dependence of $M_{h}^{\rm SpEC}$, both its amplitude and phase, displays a strong correlation with the angular projections $\left<\left(\mathcal{L}_2^\dagger S_{22}\right)^2\right|\left.S_{44}\right>$ and $  \left<S_{22}\,\mathcal{L}_1^\dagger\mathcal{L}_2^\dagger S_{22}\right|\left.S_{44}\right>$. More specifically, we observe that the four components with distinct angular dependences in Eq.~\eqref{eq:final_source_term} combine linearly in the equation. This holds true for their contributions to $M_{h}^{\rm SpEC}$ as well. The two dominant ones, $\left(\mathcal{L}_2^\dagger S_{22}\right)^2$ and $S_{22}\,\mathcal{L}_1^\dagger\mathcal{L}_2^\dagger S_{22}$, offer the major contributions to $M_{h}^{\rm SpEC}$. The comparison between the two figures suggests that a substantial portion of the spin dependence of $M_{h}^{\rm SpEC}$ originates from these two angular projections, as opposed to the radial integration. In other words, qualitatively, the angular integrals offer a means to capture the main feature of $M_{h}^{\rm SpEC}$ without the need to solve the Teukolsky equation.

This statement can also be applied to the excitation channel  $(l_L=m_L=2,n_L=0)^2 \to (l_Q=5,m_Q=4)$. In Fig.~\ref{fig:excitation_factor_220_540}, we show the corresponding $M_{h}^{\rm SpEC}$ as a function of spin. As expected, it vanishes for Schwarzschild BHs because of the selection rule in Eq.~\eqref{eq:selection_rule_schwarzschild}. Subsequently, its amplitude gradually rises to 0.00643 when the spin approaches 0.99. The overall trend is consistent with that of the angular projections in Fig.~\ref{fig:angular_projection_220_540}.

In a recent study \cite{Zhu:2024rej}, Zhu \etal extracted the value of $M_{h}^{\rm SpEC}$ for Kerr BHs by performing two types of numerical experiments: $3+1$ numerical-relativity simulations and time evolution of second-order perturbation equations. Focusing on the channel $(l_L=m_L=2,n_L=0)^2 \to (l_Q=m_Q=4)$, their results for the phase of $M_{h}^{\rm SpEC}$ show good agreements with ours (see their Fig.~1). However, the amplitude has a larger discrepancy, especially at the low spin. In particular, our amplitude for Schwarzschild BHs is 0.137,  which is 11.6\% smaller than the value of 0.153 obtained from their numerical-relativity simulations, and 27\% smaller than the value of 0.174 obtained from their time evolution of second-order perturbation equations. Additionally, the studies in Ref.~\cite{Redondo-Yuste:2023seq} yielded a smaller value $(\lesssim0.066)$ than ours. The variations across these calculations imply that the value of $M_{h}^{\rm SpEC}$ may rely on the choice of initial data in the time-domain simulations. To see this, Nakano and Ioka (hereafter NI) \cite{Nakano:2007cj} investigated the quadratic effect of Schwarzschild BHs by focusing on the even-parity perturbations. Specifically, they solved the second-order Zerilli equation sourced by the first-order Zerilli function, namely $({\rm linear\, even})\times({\rm linear\, even})\to ({\rm quadratic \, even})$. To convert their result to $M_{h}^{\rm SpEC}$, we use the following relations
\begin{equation}
\label{eq:NI_variable_Z_h}
\begin{aligned}
    &\chi^{(2)}_{\rm NI} = -2i\omega_L Z^{(2)}= -2i\omega_L\sqrt{\frac{(l_Q-2)!}{(l_Q+2)!}} h^{(2)}_{\rm SpEC}, \\
    &\chi^{(1)}_{\rm NI} = -i\omega_L Z^{(1)}= -i\omega_L\sqrt{\frac{(l_L-2)!}{(l_L+2)!}} h^{(1)}_{\rm SpEC}.
\end{aligned}
\end{equation}
Here $\chi^{(1),(2)}_{\rm NI}$ are NI's dynamical variables that appear in the first- and second-order Zerilli equations. They are related to the Zerilli functions $Z^{(1),(2)}$ through a time derivative [see the text below Eq.~(5.1) of \cite{Nakano:2007cj}]. In addition, $Z^{(1),(2)}$ can be further converted to strains $h^{(1),(2)}_{\rm SpEC}$ using the $l-$dependent coefficients in Eq.~\eqref{eq:NI_variable_Z_h}. As a result, the corresponding excitation factor for $M_{h}^{\rm SpEC}$ is given by
\begin{align}
    &M_{h}^{\rm SpEC} = \frac{\left[h^{(2)}_{\rm SpEC}\right]_{l_Q,m_Q}^{(\infty)}}{\left\{\left[h^{(1)}_{\rm SpEC}\right]_{l_L,m_L}^{(\infty)}\right\}^2}\notag \\
    &=\frac{-i\omega_L}{2}\frac{(l_L-2)!}{(l_L+2)!} \sqrt{\frac{(l_Q+2)!}{(l_Q-2)!}}\frac{\left[\chi^{(2)}_{\rm NI}\right]_{l_Q,m_Q}^{(\infty)}}{\left\{\left[\chi^{(1)}_{\rm NI}\right]_{l_L,m_L}^{(\infty)}\right\}^2} \notag \\
    &=\frac{-i\omega_L}{2}\frac{(l_L-2)!}{(l_L+2)!} \sqrt{\frac{(l_Q+2)!}{(l_Q-2)!}}M_{\chi}^{\rm NI}. 
\end{align}
As for the case of $(l_L=m_L=2,n_L=0)^2 \to (l_Q=m_Q=4)$, $M_{\chi}^{\rm NI}=0.221-0.489i$ (see TABLE II of \cite{Nakano:2007cj}), which leads to $M_{h}^{\rm SpEC}=-0.080-0.0154i$, with an amplitude of 0.081, which is smaller than our results and the numbers reported by Zhu \etal \cite{Zhu:2024rej}. A plausible source of the discrepancy lies in the composition of our linear wavefunction for $\Psi_4^{(1)}$, including both even- and odd-party contributions. The general mode coupling involves three excitation channels: $({\rm linear\, even})\times({\rm linear\, even})\to ({\rm quadratic \, even})$, $({\rm linear\, odd})\times({\rm linear\, odd})\to ({\rm quadratic \, even})$, and $({\rm linear\, odd})\times({\rm linear\, even})\to ({\rm quadratic \, odd})$. The last two contributions were omitted by NI \cite{Nakano:2007cj}, who essentially examined the quadratic effect of a pure-even mode. To our knowledge, the calculations that involve the odd parity are still missing in the literature. If the excitability for the odd sector differs from that of the even parity---despite the isospectrality of Schwarzschild BHs, the ultimate value of $M_{h}^{\rm SpEC}$ will hinge on the composite of the linear wave or, as per Zhu \etal \cite{Zhu:2024rej}, the initial data of numerical simulations. 

Finally, to close this section, we want to emphasize that the spin-weighted spheroidal harmonics used for the quadratic mode $2\omega_L$, e.g. $\tensor[_{-2}]{S}{_{44,2\omega_{220}}}$, are different from those of the linear child modes, e.g. $\tensor[_{-2}]{S}{_{44,\omega_{440}}}$, even though they have the same $l$ and $m$. This is because spin-weighted spheroidal harmonics depend on modes' complex frequencies. To compare our results for Kerr BHs with other studies, one has to collect the mode content $2\omega_L$ in all $l$'s and transform the coefficients to the new angular basis, such as spin-weighted spherical harmonics used in numerical relativity.

\section{Conclusion}
\label{sec:conclusion}
In this paper, we have developed a method to compute the excitation factor of quadratic QNMs at null infinity for Kerr BHs. The main procedure includes two essential steps (a) analytically reconstructing the metric through the CCK approach and (b) numerically solving the second-order, frequency-domain Teukolsky equation using the shooting method along a complex contour. 

We have implemented this method on two mode-coupling channels $(l_L=m_L=2,n_L=0)^2 \to (l_Q=m_Q=4)$ and $(l_L=m_L=2,n_L=0)^2 \to (l_Q=5,m_Q=4)$. The results reveal a correlation between the spin dependence of the excitation factor and the dominant angular projection, providing insights into the origin of the spin dependence. Furthermore, this feature facilitates the construction of a qualitative understanding of the spin dependence without requiring the direct solution of the Teukolsky equation. Depending on whether the angular selection rule in Eq.~\eqref{eq:selection_rule_schwarzschild} is violated or not, the excitability of the quadratic QNM increases or decreases with the BH's spin. For both cases, the factors do not vanish in the extremal limit.

Our result for the channel $(l_L=m_L=2,n_L=0)^2 \to (l_Q=m_Q=4)$ in the Schwarzschild BH case does not fully agree with the values reported by \cite{Zhu:2024rej,Redondo-Yuste:2023seq}, implying the dependence of the excitation factor on initial data (or equivalently, the parity content of the linear mode). For instance, we have shown that the excitation factor of a pure-even linear mode is different from that of a linear Teukolsky mode by comparing with results in \cite{Nakano:2007cj}. A study of couplings between even and odd modes in the future will help understand the variation caused by the ``parity freedom'' in the initial data, and construct an explicit connection between nonlinear phenomena within the Regge-Wheeler-Zerilli and Teukolsky formalism.

In the future, a valuable avenue is to generalize our calculations to the coupling between arbitrary modes and subsequently compare them with numerical-relativity simulations. The associated results will provide the theoretical foundation for assessing the detectability of the quadratic effect through various ringdown analysis methods \cite{Carullo:2019flw,Isi:2021iql,Finch:2022ynt,Ma:2023vvr,Ma:2023cwe,Wang:2023xsy}. It is also interesting to compute the excitation factor of quadratic modes near BH horizons, which has been observed by characterizing dynamical horizons in numerical relativity simulations \cite{Khera:2023oyf}. The complex contour technique will not be suitable for this calculation, and an alternative promising direction is to solve the second-order, frequency-domain Teukolsky equation on a hyperboloidal slicing. In the extremal limit, the mode coupling together with the collective excitation \cite{Yang:2013uba,Casals:2016mel} of zero-damping modes \cite{Hod:2008se,Hod:2008zz,Yang:2012he,Yang:2012pj,Yang:2013uba} may give rise to a new (nonlinear) instability in addition to the Aretakis instability \cite{Aretakis:2010gd,Aretakis:2011ha,Aretakis:2011hc,Aretakis:2012ei}. It is however also worth noting that the definition of such excitation factor at horizon may not be gauge invariant. 

Finally, as a  byproduct, we have found that the Weyl scalars $\Psi_2$ and $\Psi_3$ can be concisely expressed through the Hertz potential. These relations might be useful to study BH ringing in modified theories of gravity \cite{Li:2022pcy,Hussain:2022ins,Wagle:2023fwl}.


\begin{acknowledgments}
We thank Marc Casals for discussions during the initial stage of this work, Saul Teukolsky for pointing out the stable integration direction in Fig.~\ref{fig:cp}, and Neev Khera, Hengrui Zhu and Aaron Zimmerman for useful discussions. H. Y. is supported by the Natural Sciences and
Engineering Research Council of Canada and in part by
Perimeter Institute for Theoretical Physics. Research at Perimeter Institute is supported in part by the Government of Canada through the Department of Innovation, Science and Economic Development and by the Province of Ontario through the Ministry of Colleges and Universities.
This work makes use of the Black Hole Perturbation Toolkit.
\end{acknowledgments}

\appendix

\section{Leaver's representation}
\label{app:Leaver}
The coefficients of the three-term recurrence relation in Eq.~\eqref{eq:leaver_method_three_term_recurrence} are given as follows \cite{leaver1985analytic}
\begin{equation}
\label{eq:leaver_alpha_beta_gamma}
\begin{aligned}
    &\alpha_n=n^2+(c_0+1)n+c_0,\\
    &\beta_n=-2n^2+(2+c_1)n+c_3, \\
    &\gamma_n=n^2+(c_2-3)n+c_4-c_2+2,
\end{aligned}
\end{equation}
with
\begin{align}
    c_0=&1-s-2i\sigma_+, \notag  \\
    c_1=&-4+4i\sigma_++4ir_+\omega, \notag  \\
    c_2=&s+3-4i\omega-2i\sigma_+, \notag  \\
    c_3=&4\omega^2(2+2b)-\lambda-4am\omega-s-1\notag \\
    &+2(1+b)i\omega+(8\omega+2i)\sigma_+, \notag \\
    c_4=&s+1-(2s+1)i\omega-(8\omega+2i)\sigma_+,\notag 
\end{align}
and $b=\sqrt{1-a^2}$.

\section{Expressing spin coefficients in terms of the Hertz potential}
\label{app:spin_coefficients}
Below we list the spin coefficients in terms of the Hertz potential
\begin{alignat}{2}
    &\lambda^{(1)}= -\frac{\Delta^3}{16\bar{\Gamma}^4}\left(-\Sigma\mathcal{D}_2^\dagger\mathcal{D}_2^\dagger +2r\mathcal{D}_2^\dagger-2\right) \mathcal{D}_2^\dagger\hertzbar. \label{eq:lambda_1st_order} &\\[10pt]
    &-\frac{8\bar{\Gamma}^4}{\Delta}\bar{\sigma}^{(1)}= \Delta \Sigma  \left(-\Sigma \mathcal{D}_2\mathcal{D}^\dagger_2  - 2 + 2  \bar{\Gamma} \mathcal{D}_2 \right) \mathcal{D}^\dagger_2\hertzbar& \notag \\
    & +2i a \cos\theta\left( \Delta \Sigma \mathcal{D}^\dagger_2 - 4 a^2 r \sin^2\theta \right) \mathcal{D}^\dagger_2\hertzbar &  \notag \\
    & + 4ira \sin\theta\left(2 r -  \Sigma \mathcal{D}^\dagger_2 \right) \mathcal{L}^\dagger_2  \hertzbar.&  \\[10pt]
    &-\frac{8\sqrt{2}\bar{\Gamma}^4}{\Gamma\Delta^2}\pi^{(1)}=\bar{\Gamma}^2 \mathcal{D}_2^\dagger\left(\mathcal{D}_2^\dagger-\frac{2}{\bar{\Gamma}}\right)\mathcal{L}^\dagger_2\hertzbar &\notag \\
    & +\frac{ia\sin\theta}{\Gamma} (2\Gamma - 2\Sigma \mathcal{D}^\dagger_2 +  \bar{\Gamma}^2 \mathcal{D}^\dagger_2)\mathcal{D}^\dagger_2\hertzbar.&\\[10pt]
    &\frac{8\sqrt{2}\bar{\Gamma}^2}{\Gamma\Delta^2}\bar{\tau}^{(1)}=\left(\mathcal{D}_2^\dagger+\frac{1}{\Gamma}-\frac{1}{\bar{\Gamma}}\right) \left(\mathcal{D}_2^\dagger-\frac{1}{\Gamma}-\frac{1}{\bar{\Gamma}}\right) \mathcal{L}^\dagger_2\hertzbar &\notag \\
    & -\frac{ia\sin\theta}{\Sigma^2} \left[2 ( \Sigma-\Gamma^2 + \bar{\Gamma}^2)-\Sigma  (2\bar{\Gamma}-\Gamma) \mathcal{D}^\dagger_2\right]\mathcal{D}^\dagger_2\hertzbar.&
\end{alignat}   
During the calculation, we find that it is convenient to introduce the following two variables:
\begin{align}
    &\alpha_w^{(1)}
    \equiv\alpha^{(1)}-\frac{\beta^{(0)}}{2}h_{\bar{m}\bar{m}}, \quad \bar{\beta}_w^{(1)}\equiv\bar{\beta}^{(1)}+\frac{\beta^{(0)}}{2}h_{\bar{m}\bar{m}}. \label{eq:alphaw_betaw} 
\end{align}
Their expressions are given by 
\begin{align}
    &-\frac{8\sqrt{2}\bar{\Gamma}^3}{\Delta^2}\alpha_w^{(1)}=[-1 - r \mathcal{D}^\dagger_0 + \Sigma \mathcal{D}^\dagger_1\mathcal{D}^\dagger_2] \mathcal{L}_2^\dagger\hertzbar \notag \\
    &+i a \cos\theta \mathcal{D}^\dagger_2 \mathcal{L}^\dagger_2\hertzbar + 3i a \sin\theta \mathcal{D}^\dagger_2\hertzbar \notag \\
    & + 2 a^2  \sin\theta \cos\theta\mathcal{D}^\dagger_1\mathcal{D}^\dagger_2\hertzbar - 2ir a \sin\theta \mathcal{D}^\dagger_2\mathcal{D}^\dagger_2\hertzbar. \\
    &-\frac{16\sqrt{2}\bar{\Gamma}^3\Gamma}{\Delta^2}\bar{\beta}_w^{(1)}= [-6\Gamma - 2r\Gamma\mathcal{D}^\dagger_0 + (\Gamma^2 + 7\Sigma)\mathcal{D}^\dagger_2   \notag \\
    & - \frac{4\Gamma\Sigma(r-1)}{\Delta}\mathcal{D}^\dagger_2 - 4\bar{\Gamma}]\mathcal{L}_2^\dagger\hertzbar \notag \\
    &+ia\sin\theta (-6\Gamma - 2 \Gamma^2\mathcal{D}^\dagger_0+ 4r \Gamma\mathcal{D}^\dagger_1 + 4\bar{\Gamma})\mathcal{D}^\dagger_2\hertzbar.
\end{align}
The expression of $\epsilon^{(1)}$, $\rho^{(1)}$, and $\kappa^{(1)}$ are quite lengthy. They can be constructed from four building blocks:
\begin{align}
    &(\bm{\delta}+2\beta+2\tau-\pi)^{(0)}h_{l\bar{m}}=-\frac{\Delta}{4\bar{\Gamma}^3} \left[ \bar{\Gamma}(\Sigma\mathcal{D}_2^\dagger-2r)\mathcal{L}_1^\dagger\mathcal{L}_2^\dagger\right.\notag \\
    &\left.+4ia\sin\theta (2r-\Sigma\mathcal{D}_2^\dagger)\mathcal{L}_2^\dagger +2a^2\sin^2\theta (\bar{\Gamma}-2\Gamma)\mathcal{D}_2^\dagger\right]\hertzbar. \label{eq:app:hlmbar} \\[10pt]
    &(-\bm{\Delta}+2\bar{\gamma}+\mu-\bar{\mu})^{(0)}h_{ll}=-\frac{\Delta}{4\bar{\Gamma}^2}[(\Gamma-\bar{\Gamma}-\Sigma\mathcal{D}_1^\dagger)\mathcal{L}_1^\dagger\notag \\
    &+2ia\sin\theta(1-\bar{\Gamma}\mathcal{D}_1^\dagger)]\mathcal{L}_2^\dagger\hertzbar + c.c.\,. \\[10pt]
    &(\bm{D}-\rho)^{(0)}h_{l\bar{m}}=-\frac{\Delta}{2\sqrt{2}\bar{\Gamma}^3}[(\Sigma^2\mathcal{D}_1\mathcal{D}_2^\dagger-2r\Sigma\mathcal{D}_1+2\bar{\Gamma}\Sigma\mathcal{D}_2^\dagger \notag \\
    &-2\Sigma+\Gamma^2-\bar{\Gamma}^2)\mathcal{L}_2^\dagger -4ia^3\cos^2\theta\sin\theta\mathcal{D}_2^\dagger \notag \\
    &+a^2\Sigma\sin2\theta \mathcal{D}_1\mathcal{D}_2^\dagger]\hertzbar. \\[10pt]
    &2\sqrt{2}\bar{\Gamma}\bm{\delta}^{(0)}h_{ll}= [2 a^2 \sin ^2\theta +2 i a \sin\theta \Gamma (\mathcal{L}_1-\mathcal{L}_1^\dagger)  \notag \\
    &+\Gamma^2 \mathcal{L}_0\mathcal{L}_1^\dagger]\mathcal{L}_2^\dagger\hertzbar+(2 a^2 \sin ^2\theta +\bar{\Gamma}^2 \bar{\mathcal{L}}_0\bar{\mathcal{L}}_1^\dagger)\bar{\mathcal{L}}_2^\dagger\hertz. \label{eq:app:hll}
\end{align}
Combining with Eq.~\eqref{eq:app:hlmbar}--\eqref{eq:app:hll}, \eqref{eq:NP_spin_coefficients} and \eqref{eq:reconstruction_all_h}, one can use the following relations to obtain their expressions
\begin{align}
    \epsilon^{(1)}=&\frac{1}{4}(-\bm{\Delta}+2\bar{\gamma}+\mu-\bar{\mu})^{(0)}h_{ll} \notag \\
    &+\frac{1}{4}(-\bm{\delta}-2\beta+\bar{\pi}-2\tau)^{(0)}h_{l\bar{m}}\notag \\
    &+\frac{1}{4}(\bm{\bar{\delta}}+2\bar{\beta}-5\pi-2\bar{\tau})^{(0)}h_{lm}. \\
    \rho^{(1)}=&-\frac{1}{2}(\bm{\delta}+2\beta-\bar{\pi}+2\tau)^{(0)}h_{l\bar{m}}\notag \\
    &+\frac{1}{2}(\bm{\bar{\delta}}-3\pi+2\bar{\beta})^{(0)}h_{lm}+\frac{1}{2}\mu^{(0)} h_{ll}. \\
    \kappa^{(1)}=&(\bm{D}-\bar{\rho})^{(0)}h_{lm}-\frac{1}{2}(\bm{\delta}-\bar{\pi}+\tau)^{(0)}h_{ll}.
\end{align}
Note that the expressions above have been simplified in ORG. See e.g., \cite{Loutrel:2020wbw} for the full expressions.

For Schwarzschild BHs, the results can be simplified to
\begin{align}
    &\lambda^{(1)}= \frac{\Delta^3}{16r^2}\left(\mathcal{D}_2^\dagger -\frac{1}{r}\right)^2 \mathcal{D}_2^\dagger\hertzbar. \\[10pt]
    &\bar{\sigma}^{(1)}=\frac{\Delta^2}{8}\mathcal{D}_2\left(\mathcal{D}_2^\dagger -\frac{2}{r}\right) \mathcal{D}_2^\dagger\hertzbar. \\[10pt]
    &\bar{\kappa}^{(1)} =  \frac{r\Delta}{2\sqrt{2}} 
    \left(\mathcal{D}_1+\frac{2}{r}\right)\left(\mathcal{D}_2^\dagger -\frac{2}{r}\right)\eth\hertzbar \notag \\
    &+ \frac{r}{4\sqrt{2}}\bar{\eth}\eth\eth\hertzbar+\frac{r}{4\sqrt{2}}\bar{\eth}\bar{\eth}\bar{\eth}\hertz. \\[10pt]
    &\epsilon^{(1)}=-\frac{1}{8}\bar{\eth}\bar{\eth}\hertz+\frac{\Delta}{8} (\mathcal{D}_1^\dagger+\frac{1}{\Delta})\eth\eth\hertzbar. \\[10pt]
   &\rho^{(1)} = -\frac{\Delta}{8}\left(\bar{\mathcal{D}}_2^\dagger -\frac{1}{r}\right)\bar{\eth}\bar{\eth}\hertz+\frac{\Delta}{8}\left(\mathcal{D}_2^\dagger -\frac{3}{r}\right)\eth\eth\hertzbar. \\[10pt]
   &\bar{\tau}^{(1)}=-\frac{\Delta^2}{8\sqrt{2}r}\left(\mathcal{D}_2^\dagger -\frac{1}{r}\right)^2\eth\hertzbar. \\[10pt]
   &\pi^{(1)}=\frac{\Delta^2}{8\sqrt{2}r}\left(\mathcal{D}_2^\dagger -\frac{1}{r}\right)^2\eth\hertzbar. \\
    &\alpha_w^{(1)}=\frac{\Delta^2}{8\sqrt{2}r}\left(\mathcal{D}_1^\dagger\mathcal{D}_2^\dagger-\frac{\mathcal{D}_0^\dagger}{r}-\frac{1}{r^2}\right)\eth\hertzbar. \\
    &\bar{\beta}_w^{(1)}=\frac{\Delta^2}{8\sqrt{2}r}\left(-\frac{2}{\Delta}\mathcal{D}^\dagger_2+ \frac{\mathcal{D}^\dagger_2}{r} +\frac{4}{\Delta r}- \frac{1}{r^2}
\right)\eth\hertzbar.
\end{align}
Notice that $\alpha_w^{(1)}$ and $\bar{\beta}_w^{(1)}$ have a certain spin weight.

\section{\texorpdfstring{Expressing $\Psi_2^{(1)}$ in terms of the Hertz potential}{Derivation of psi2}}
\label{app:psi2_hertz}
We start with the Ricci identity \cite{chandrasekhar1998mathematical}.
\begin{align}
    3\Psi_2&=(\bm{\bar{\delta}}-2\alpha+\bar{\beta}-\pi-\bar{\tau})\beta-(\bm{\delta}-\bar{\alpha}+\bar{\pi}+\tau)\alpha \notag \\
    &+(\bm{D}+\epsilon+\bar{\epsilon}+\rho-\bar{\rho})\gamma-(\bm{\Delta}-\bar{\gamma}-\gamma+\bar{\mu}-\mu)\epsilon \notag \\
    &+(\bm{\bar{\delta}}-\alpha+\bar{\beta}-\bar{\tau}-\pi)\tau-(\bm{\Delta}-\bar{\gamma}-\gamma+\bar{\mu}-\mu)\rho \notag \\
    &+2(\nu\kappa-\lambda\sigma).
\end{align}
To the linear order, the identity simplifies to 
\begin{align}
    &3\Psi_2^{(1)}=(\bm{\bar{\delta}}-3\alpha+\bar{\beta}-\pi-\bar{\tau})^{(1)}\beta^{(0)} -(\bm{\delta}+2\tau)^{(0)}\alpha^{(1)} \notag \\
    & +(\bm{\bar{\delta}}+\bar{\beta}-\bar{\tau}-\pi)^{(1)}\tau^{(0)} \notag \\
    &+(\bm{D}+\epsilon+\bar{\epsilon}+\rho-\bar{\rho})^{(1)}\gamma^{(0)}-(\bm{\Delta} -2\gamma)^{(0)}(\epsilon+\rho)^{(1)}\notag \\
    &+(\bm{\bar{\delta}}-2\alpha+\bar{\beta}-\pi-\bar{\tau})^{(0)}\beta^{(1)}-(\bm{\delta}-\bar{\alpha}+\bar{\pi}+\tau)^{(1)}\alpha^{(0)}\notag \\
    &+(\bm{\bar{\delta}}-\alpha+\bar{\beta}-\bar{\tau}-\pi)^{(0)}\tau^{(1)},  \label{eq:linearized_psi2_middle}
\end{align}
where we have used the facts that $\lambda^{(0)}=\sigma^{(0)}=\nu^{(0)}=\nu^{(1)}=\gamma^{(1)}=\mu^{(1)}=\bm{\Delta}^{(1)}=0$ and $\alpha^{(0)}=\pi^{(0)}-\bar{\beta}^{(0)}$.

\subsection{\texorpdfstring{The first line of Eq.~\eqref{eq:linearized_psi2_middle}}{linearized_psi2_middle}}
Many terms in the first line of Eq.~\eqref{eq:linearized_psi2_middle} can be elegantly canceled. To see this, we use the following identities \cite{Ripley:2020xby}
\begin{align}
    &\bar{\beta}_w^{(1)}-\alpha^{(1)}_w+\bar{\mu}^{(0)}h_{l\bar{m}}+\frac{\bar{\pi}^{(0)}}{2}h_{\bar{m}\bar{m}}=\frac{1}{2}(\bm{\delta}+4\beta)^{(0)} h_{\bar{m}\bar{m}}, \label{eq:identities_beta_alpha} \\ 
    &\bm{\Delta}^{(0)}\beta^{(0)}+\bar{\mu}^{(0)}\beta^{(0)}=0, \label{eq:identities_Delta_beta} \\
    & \pi^{(1)}+\bar{\tau}^{(1)}=-\frac{\bar{\pi}^{(0)}+\tau^{(0)}}{2}h_{\bar{m}\bar{m}}. \label{eq:identities_pi_tau}
\end{align}
We first adopt Eq.~\eqref{eq:identities_pi_tau} and the complex conjugate of Eq.~\eqref{eq:directional_derivatives_delta1}, which leads to
\begin{align}
&(\bm{\bar{\delta}}-3\alpha+\bar{\beta}-\pi-\bar{\tau})^{(1)}\beta^{(0)} -(\bm{\delta}+2\tau)^{(0)}\alpha^{(1)}  \notag \\
&=(-h_{l\bar{m}}\bm{\Delta}^{(0)}+\frac{1}{2}h_{\bar{m}\bar{m}}\bm{\delta}^{(0)}-\alpha^{(1)}+\bar{\beta}^{(1)})\beta^{(0)} \notag \\
&+\frac{\bar{\pi}^{(0)}+\tau^{(0)}}{2}h_{\bar{m}\bar{m}}\beta^{(0)}-(\bm{\delta}^{(0)}+2\beta^{(0)})\alpha^{(1)}-2\tau^{(0)}\alpha^{(1)}.
\end{align}
Then we use Eq.~\eqref{eq:alphaw_betaw} to replace some $\alpha^{(1)}$ and $\bar{\beta}^{(1)}$ with $\alpha_w^{(1)}$ and $\bar{\beta}_w^{(1)}$. This yields
\begin{align}
    &{\rm{L.H.S}}=(-h_{l\bar{m}}\bm{\Delta}^{(0)}-\bar{\mu}h_{l\bar{m}})\beta^{(0)} \notag \\
    &+\frac{\tau^{(0)}}{2}h_{\bar{m}\bar{m}}\beta^{(0)}-(\bm{\delta}^{(0)}+2\beta^{(0)})\alpha^{(1)}_w-2\tau^{(0)}\alpha^{(1)}, \label{app:eq:first_line_middle_result}
\end{align}
where we have used Eq.~\eqref{eq:identities_beta_alpha}. Here L.H.S refers to the first line of Eq.~\eqref{eq:linearized_psi2_middle}. Finally, the first line of Eq.~\eqref{app:eq:first_line_middle_result} vanishes due to Eq.~\eqref{eq:identities_Delta_beta}, and we end up with
\begin{align}
    &{\rm{L.H.S}}= -(\bm{\delta}+2\beta)^{(0)}\alpha_w^{(1)} -2\tau^{(0)} \alpha_w^{(1)}-\frac{\tau^{(0)}}{2}h_{\bar{m}\bar{m}}\beta^{(0)} \notag \\
    &=-\frac{\tau^{(0)}}{2}h_{\bar{m}\bar{m}}\beta^{(0)}-\frac{1}{\sqrt{2}\Gamma}\left(\mathcal{L}_1^\dagger -\frac{2ia\sin\theta}{\bar{\Gamma}}\right) \alpha_w^{(1)}. \label{eq:psi2_sim_a}
\end{align}

\subsection{\texorpdfstring{The second line of Eq.~\eqref{eq:linearized_psi2_middle}}{linearized_psi2_middle}}
By using 
\begin{subequations}
\begin{align}
    &\bm{\delta}^{(0)} \tau^{(0)}=\tau^{(0)} (\tau+2\beta-\bar{\pi})^{(0)}, \\
    &\bm{\Delta}^{(0)} \tau^{(0)} = -\tau^{(0)}(\mu+\bar{\mu})^{(0)},
\end{align}  
\end{subequations}
and Eq.~\eqref{eq:identities_pi_tau},
the second line of Eq.~\eqref{eq:linearized_psi2_middle} can be reduced to 
\begin{align}
    &(\bm{\bar{\delta}}+\bar{\beta}-\bar{\tau}-\pi)^{(1)}\tau^{(0)}=\frac{\tau^{(0)}}{2}h_{\bar{m}\bar{m}}\beta^{(0)}\notag \\
    &+\left[h_{l\bar{m}}(\mu^{(0)}+\bar{\mu}^{(0)})+h_{\bar{m}\bar{m}}\tau^{(0)}+\bar{\beta}^{(1)}_w\right]\tau^{(0)} . \label{eq:psi2_sim_b}
\end{align}
Note that the first term of Eqs.~\eqref{eq:psi2_sim_a} and \eqref{eq:psi2_sim_b} are identical, resulting in additional cancellation.

\subsection{\texorpdfstring{The last two lines of Eq.~\eqref{eq:linearized_psi2_middle}}{linearized_psi2_middle}}
In Appendix \ref{app:psi2_sim_c}, we show that
\begin{align}
    &(\bm{\bar{\delta}}-2\alpha+\bar{\beta}-\pi-\bar{\tau})^{(0)}\beta^{(1)}-(\bm{\delta}-\bar{\alpha}+\bar{\pi}+\tau)^{(1)}\alpha^{(0)} \notag \\
    &=(\bm{\bar{\delta}}+2\bar{\beta}-\pi-\bar{\tau})^{(0)}\beta_w^{(1)}-\bar{\alpha}_w^{(1)}\pi^{(0)}\notag \\
    &-\pi^{(0)}\left(\bm{\bar{\delta}}+2\bar{\beta}-\frac{5\pi+\bar{\tau}}{2}\right)^{(0)}h_{mm}. \label{eq:psi2_sim_c}
\end{align}
After inserting the result above, along with the expression for $h_{mm}$ in Eq.~\eqref{eq:reconstruction_hmbarmbar} and spin coefficients in Appendix \ref{app:spin_coefficients}, into the last two lines of Eq.~\eqref{eq:linearized_psi2_middle}, we find it vanishes identically, namely
\begin{align}
    &(\bm{\bar{\delta}}-2\alpha+\bar{\beta}-\pi-\bar{\tau})^{(0)}\beta^{(1)}-(\bm{\delta}-\bar{\alpha}+\bar{\pi}+\tau)^{(1)}\alpha^{(0)}\notag \\
    &+(\bm{\bar{\delta}}-\alpha+\bar{\beta}-\bar{\tau}-\pi)^{(0)}\tau^{(1)}=0.  \label{eq:psi2_sim_c_final}
\end{align}

\subsection{Final result}
Finally, by collecting the outcomes in Eqs.~\eqref{eq:psi2_sim_a}, \eqref{eq:psi2_sim_b}, and \eqref{eq:psi2_sim_c_final} along with the relations 
\begin{subequations}
\begin{align}
&\left[\mathcal{D}_{m}^\dagger,\mathcal{D}_{n}^\dagger\right]=(n-m)\frac{2}{\Delta}\left[1-\frac{2(r-1)^2}{\Delta}\right], \label{eq:chandra_D_commutator} \\
&\mathcal{D}_3^\dagger\mathcal{D}_1^\dagger-\mathcal{D}_2^\dagger\mathcal{D}_2^\dagger=-\frac{2}{\Delta},  
\end{align}
\end{subequations}
we arrive at our final result for $\Psi_2^{(1)}$ in Eq.~\eqref{eq:psi2_hertz_final}.

\section{\texorpdfstring{Derivation of Eq.~\eqref{eq:psi2_sim_c}}{Derivation of Equationc}}
\label{app:psi2_sim_c}
Replacing $\alpha^{(0)}$ with $\pi^{(0)}-\bar{\beta}^{(0)}$ yields two terms
\begin{align}
    &(\bm{\bar{\delta}}-2\alpha+\bar{\beta}-\pi-\bar{\tau})^{(0)}\beta^{(1)}+(\bm{\delta}-\bar{\alpha}+\bar{\pi}+\tau)^{(1)}\bar{\beta}^{(0)}, \label{eq:psi2_sim_c1}
\end{align}
and 
\begin{align}
    -(\bm{\delta}-\bar{\alpha}+\bar{\pi}+\tau)^{(1)}\pi^{(0)}. \label{eq:psi2_sim_c2}
\end{align}

\subsection{\texorpdfstring{Eq.~\eqref{eq:psi2_sim_c1}}{Derivation of Equationc1}}
Replacing $\beta^{(1)}$ with $\beta_w^{(1)}$, the first part of Eq.~\eqref{eq:psi2_sim_c1} becomes
\begin{align}
    &(\bm{\bar{\delta}}+2\bar{\beta}-3\pi-\bar{\tau})^{(0)}\beta^{(1)} \notag \\
    &=(\bm{\bar{\delta}}+2\bar{\beta}-3\pi-\bar{\tau})^{(0)}\beta_w^{(1)}-\frac{\bar{\beta}^{(0)}}{2}(2\bar{\beta}-3\pi-\bar{\tau})^{(0)}h_{mm} \notag \\
    &
    -\frac{h_{mm}}{2}\bm{\bar{\delta}}^{(0)}\bar{\beta}^{(0)}    -\frac{\bar{\beta}^{(0)}}{2}\bm{\bar{\delta}}^{(0)}h_{mm}. \label{app:eq:psi2_retrograde_first}
\end{align}
Meanwhile, the second part of Eq.~\eqref{eq:psi2_sim_c1} can be converted to
\begin{align}
&(\bm{\delta}+\beta-\bar{\alpha}+\bar{\pi}+\tau)^{(1)}\bar{\beta}^{(0)}  \notag \\
&=\left(-h_{lm}\bm{\Delta}^{(0)}+\frac{1}{2}h_{mm}\bm{\bar{\delta}}^{(0)}+\beta-\bar{\alpha}+\bar{\pi}+\tau\right)^{(1)}\bar{\beta}^{(0)} \notag \\
&=\left(h_{lm}\mu^{(0)}+\frac{1}{2}h_{mm}\bm{\bar{\delta}}^{(0)}+\beta^{(1)}-\bar{\alpha}^{(1)}-\frac{\pi^{(0)}+\bar{\tau}^{(0)}}{2}h_{mm}\right)\bar{\beta}^{(0)} \notag \\
&=\left(h_{lm}\mu^{(0)}+\beta_w^{(1)}-2\bar{\beta}^{(0)}h_{mm}-\bar{\alpha}_w^{(1)}-\frac{\pi^{(0)}+\bar{\tau}^{(0)}}{2}h_{mm}\right)\bar{\beta}^{(0)} \notag \\
&+\frac{1}{2}h_{mm}\bm{\bar{\delta}}^{(0)}\bar{\beta}^{(0)}+\bar{\beta}^{(0)\,2}h_{mm}, \label{app:eq:psi2_retrograde_second}
\end{align}
where we have used Eq.~\eqref{eq:directional_derivatives_delta1} for the first equality, Eq.~\eqref{eq:identities_Delta_beta} for the second equality, and Eq.~\eqref{eq:alphaw_betaw} for the last one.

Combining Eqs.~\eqref{app:eq:psi2_retrograde_first} and \eqref{app:eq:psi2_retrograde_second} yields 
\begin{align}
    {\rm{Eq.~\eqref{eq:psi2_sim_c1}}}&=(\bm{\bar{\delta}}+2\bar{\beta}-3\pi-\bar{\tau})^{(0)}\beta_w^{(1)}+\frac{\bar{\beta}^{(0)}\pi^{(0)}}{2}h_{mm} \notag \\
    &+\left(\beta_w^{(1)}-\bar{\alpha}_w^{(1)}+\mu^{(0)}h_{lm}+\frac{\pi^{(0)}}{2}h_{mm}\right.\notag \\
    &\left.-2\bar{\beta}^{(0)}h_{mm}-\frac{1}{2}\bm{\bar{\delta}}^{(0)}h_{mm}\right)\bar{\beta}^{(0)}.
\end{align}
The last parenthesis vanishes because of Eq.~\eqref{eq:identities_beta_alpha}. Then
\begin{align}
    &{\rm{Eq.~\eqref{eq:psi2_sim_c1}}}=\frac{\bar{\beta}^{(0)}\pi^{(0)}}{2}h_{mm}+(\bm{\bar{\delta}}+2\bar{\beta}-3\pi-\bar{\tau})^{(0)}\beta_w^{(1)}. \label{eq:psi2_sim_c1_final}
\end{align}

\subsection{\texorpdfstring{Eq.~\eqref{eq:psi2_sim_c2}}{Derivation of Equationc1}}
With the help of
\begin{align}
    \bm{\delta}^{(0)}\bar{\pi}^{(0)}=2\bar{\pi}(\beta-\bar{\pi}),\quad \bm{\Delta}^{(0)}\bar{\pi}^{(0)}=-2\bar{\pi}^{(0)}\bar{\mu}^{(0)},
\end{align}
it is straightforward to obtain
\begin{align}
    &{\rm{Eq.~\eqref{eq:psi2_sim_c2}}}=-\frac{\bar{\beta}^{(0)}\pi^{(0)}}{2}h_{mm}\notag \\
    &-\left(2h_{lm}\mu^{(0)}-\bar{\alpha}_w^{(1)}-\frac{3\pi^{(0)}+\bar{\tau}^{(0)}}{2}h_{mm}\right)\pi^{(0)}. \label{eq:psi2_sim_c2_final}
\end{align}

\subsection{Final result}
Finally, we combine Eq.~\eqref{eq:psi2_sim_c1_final} and Eq.~\eqref{eq:psi2_sim_c2_final} to obtain Eq.~\eqref{eq:psi2_sim_c} in the main text. Notice that the first term of Eq.~\eqref{eq:psi2_sim_c1_final} and Eq.~\eqref{eq:psi2_sim_c2_final} cancel each other.

\section{\texorpdfstring{The source of the second-order Teukolsky equation for Schwarzschild BHs with $l_L=2$ and $l_Q=4$}{Schwarzschild l2}}
\label{app:source_Schwarzschild_l2}
The source $Q(r)$ in Eq.~\eqref{eq:2nd_equation_with_angular_projection} reads
\begin{widetext}
\begin{align}
   &\frac{48\sqrt{7\pi}}{5} Q(r)= r^2\left[-13 r^{10} \omega ^6-34 i r^9 \omega ^5+r^8 \left(35+269 i \omega
   \right)\omega^4+r^7 \left(131 i\omega+99 \right)i\omega ^3+r^6 \left(865 \omega ^2-949
   i \omega+44\right)\omega ^2 \right.\notag \\
   &+r^5 \left(1249 \omega ^2+530 i\omega-35
   \right)i\omega+r^4 \left(-53 \omega ^2-196i \omega+557 \right)i\omega+r^3 \left(2144
   \omega ^2-1540 i \omega +120\right)+r^2 \left(-1380 \omega ^2+930 i \omega
   -564\right) \notag \\
   &\left.+r (816+408 i \omega )-168 i \omega -336\right]\times\left(\tensor[_{+2}]{{R_{22\omega}}}{}\right)^2 \notag \\
   &+ r^2\Delta\left[32 i r^8 \omega ^5-78 r^7 \omega ^4-r^6 \left(424 i\omega+130\right)i\omega ^3+r^5
   \left(154+582 i \omega\right)\omega ^2+r^4 \left(-968  \omega ^2+1227 i\omega
   +42  \right) i\omega-r^3 \left(2320 i\omega+300  \right)i\omega\right. \notag \\
   &\left.+r^2 \left(-936
   \omega ^2+1291 i \omega +48\right)-r (192+2352 i \omega )+1152 i \omega +192\right]\times \tensor[_{+2}]{{R_{22\omega}}}{}\left(\frac{d}{dr}\tensor[_{+2}]{{R_{22\omega}}}{}\right)\notag \\
    &+\Delta^2[19 r^8 \omega ^4+50 i r^7 \omega ^3-r^6 \left(86+143 i \omega \right)\omega ^2-r^5
   \left(339i \omega+91 \right)i \omega +r^4 \left(-339 \omega ^2+538 i \omega
   +42\right)-r^3 (306+945 i \omega )\notag \\
   &+r^2 (810+447 i \omega )-912 r+360]\times\left(\frac{d}{dr}\tensor[_{+2}]{{R_{22\omega}}}{}\right)^2. 
\end{align} 
\end{widetext}
Here $\omega$ stands for the linear mode $\omega_L$.
The structure of $Q(r)$ follows Eq.~\eqref{eq:leaver_source_q_R_dr_R} and its asymptotic expansion is given in Eq.~\eqref{eq:Q_asymptotic_expansion}.

\def\bibsection{\section*{References}}
\bibliography{References}

\end{document}